\documentclass[10pt,superscriptaddress,aps,pra,onecolumn,showpacs,floatfix]{revtex4-2}
\usepackage[utf8]{inputenc}
\usepackage{graphicx,amsmath,amsfonts,amssymb}
\usepackage{color}
\usepackage[colorlinks=true, allcolors={blue}]{hyperref}
\usepackage{graphicx}
\usepackage{epstopdf}
\usepackage{float}
\usepackage{placeins}
\usepackage{lipsum}
\usepackage{comment}

\usepackage{soul}

\usepackage{silence}
\usepackage{xcolor}

\usepackage[normalem]{ulem}
\usepackage{orcidlink}

\begin{document}

\preprint{APS/123-QED}

\title{Multiparameter estimation with position-momentum correlated Gaussian probes}

\author{João C. P. Porto \orcidlink{0009-0006-6639-1413}}
%\email{carlosciaufpi@gmail.com}
\affiliation{Departamento de F\'{i}sica, Universidade Federal do Piau\'{i}, Campus Ministro Petr\^{o}nio Portela, CEP 64049-550, Teresina, PI, Brazil}

\author{Carlos H. S. Vieira~\orcidlink{0000-0001-7809-6215}}
%\email{carloshsv09@gmail.com}
\affiliation{Centro de Ci\^{e}ncias Naturais e Humanas, Universidade Federal do ABC,
Avenida dos Estados 5001, 09210-580 Santo Andr\'e, S\~{a}o Paulo, Brazil.}

\author{Pedro R. Dieguez \orcidlink{0000-0002-8286-2645}}
%\email{Corresponding author: pedro.dieguez@ug.edu.pl}
\affiliation{International Centre for Theory of Quantum Technologies, University of Gdańsk, Jana Bażyńskiego 1A, 80-309 Gdańsk, Poland}

\author{Irismar G. da Paz~\orcidlink{0000-0002-9613-9642}}
%\email{email@gmail.com}
\affiliation{Departamento de F\'{i}sica, Universidade Federal do Piau\'{i}, Campus Ministro Petr\^{o}nio Portela, CEP 64049-550, Teresina, PI, Brazil}

\author{Lucas S. Marinho \orcidlink{0000-0002-2923-587X}}
\email{Corresponding author: lucas.marinho@ufpi.edu.br}
\affiliation{Departamento de F\'{i}sica, Universidade Federal do Piau\'{i}, Campus Ministro Petr\^{o}nio Portela, CEP 64049-550, Teresina, PI, Brazil}

\date{\today}% It is always \today, today,

\begin{abstract}
Gaussian quantum probes have been widely used in quantum metrology and thermometry, where the goal is to estimate the temperature of an environment with which the probe interacts. It was recently shown that introducing initial position-momentum (PM) correlations in such probes can enhance the estimation precision compared to standard, uncorrelated Gaussian states. Motivated by these findings, we investigate whether PM correlations can also be advantageous in a simultaneous estimation setting, specifically, when estimating both the PM correlations themselves and the effective environment temperature that interacts with the probe. Using the Quantum Fisher Information Matrix, we derive new precision bounds for this joint estimation task. Additionally, we demonstrate that such correlations can serve as a resource to improve temperature estimation within this multiparameter context. Finally, we analyze the compatibility between the two parameters, establishing conditions under which the derived bounds can be saturated.
\end{abstract}

\maketitle

%\tableofcontents

\section{Introduction}\label{sec:intro}

Quantum metrology seeks to devise optimal strategies utilizing quantum resources to achieve ultra-precise and highly sensitive measurements of physical quantities, surpassing the limits imposed by classical physics \cite{GiovannettiScience2004,Giovannetti2011Nature}. Its applications include improving time and frequency
standards \cite{Katori2011NatPhotonics}, superresolution imaging~\cite{BraunPRA2023}, magnetic field detection for biomedical diagnostics \cite{Taylor2016PhysRep}, ultra-precise clocks~\cite{RosenbandSCIENCE2008}, navigation systems~\cite{GracePhysRevApplied2020}, magnetometry~\cite{Budker2007NATURE}, optical and gravitational-wave interferometry~\cite{DEMKOWICZDOBRZANSKI2015,TsePRL2019}, and thermometry~\cite{WengPRL2014,WuPRR2021,AnPRApplied2022,LandiPRA2024}. Although most existing studies have concentrated on the estimation of individual parameters~\cite{AdessoThermometryPRL2015,MonrasPRA2006,MonrasPRL2007}, the simultaneous estimation of joint parameters remains relatively underexplored  (some contributions are~\cite{Monras_IlluminatiPRA2011,Knight_AdessoPRA2013,monras2013ARXIV,PinelPRA2013,Nichols_AdessoPRA2018}). 
In general, estimating $\kappa$ parameters individually often requires $\kappa$ times more resources, such as energy, coherence, or entanglement during preparation of the input state~\cite{Braun_PirandolaRevModPhys2018, RagyPRA2016,Nichols_AdessoPRA2018}, compared to simultaneous estimation in a single experimental run. However, the advantage of simultaneous estimation is not always ensured, especially when the parameters involved are not mutually \textit{compatible}. The parameters are considered \textit{compatible} when the following conditions are met~\cite{Braun_PirandolaRevModPhys2018}: (i) the existence of a single probe state that allows optimal sensitivity for all parameters of interest, (ii) the existence of a single measurement capable of optimally extracting information about all parameters from the probe state, and (iii) the statistical independence of the estimated parameters. When these conditions are not met, quantum advantages in multiparameter estimation may be fundamentally constrained by the non-commuting nature of quantum measurements~\cite{Szczykulska2016Review}.

A wide range of advanced applications inherently involves the estimation of multiple parameters. Examples include quantum-enhanced imaging techniques~\cite{TsangPRA2017}, three-dimensional magnetic field sensing~\cite{Baumgratz_DattaPRL2016}, high-precision characterization of components crucial for fault-tolerant quantum technologies~\cite{martinis2015qubitmetrologybuildingfaulttolerant}, and quantum information science~\cite{PreskillJMordOpt2000}. Moving beyond single-parameter estimation, recent work has shown that impurity probes can be effectively employed for multiparameter estimation at finite temperatures~\cite{Mihailescu_2024QST}. Quantum metrology has also been applied to systems where two-level atoms interact with a single-mode electromagnetic field, enabling the simultaneous estimation of important quantities such as the photon number inside a cavity and the detuning parameter~\cite{Houssaoui2023EPJP}. Furthermore, it has been demonstrated that, for a fixed average energy, the joint estimation of multiple phases can yield higher overall precision than estimating each phase separately~\cite{DattaPRA2016}. Similarly, when operating under a fixed resource budget, simultaneous quantum estimation of parameters linked to non-commuting unitary generators has been shown to outperform individual estimation strategies~\cite{WalmsleyPRL2013,BaumgratzPRL2016}.

Recent work has demonstrated that initial position-momentum (PM) correlations in Gaussian states can enhance precision in single-parameter quantum estimation tasks~\cite{Porto2024,porto2025NonMarkovian}. In line with this advantage, here we investigate whether similar benefits can be achieved in a multiparameter scenario by employing a PM-correlated Gaussian state as a quantum probe. In our approach, we introduce a real-valued parameter that controls the degree of such correlation. When this parameter is set to zero, the state reduces to the standard uncorrelated Gaussian wave packet. PM correlations have long been studied through the quantized operators $\hat{x}$ and $\hat{p}$, where $[\hat{x}, \hat{p}] = i\hbar$~\cite{bohm1951quantum}. These correlated states not only saturate the Robertson–Schrödinger uncertainty relation, but also serve as a generalization of Glauber’s coherent states~\cite{DODONOV1980PLA,Glauber1963}. Physically, such states can be generated by the propagation of atomic beams through a transverse harmonic potential, which acts as a focusing lens and imprints a quadratic phase on the wave function~\cite{Janicke1995}. PM-correlated Gaussian wave packets have found applications in quantum coherence~\cite{PP2023,PP2024}, quantum optics and matter-wave interferometry~\cite{Campos1999JMO,OzielMPLA2019,LustosaPRA2020,MarinhoPRA2020,Marinho_2023,Marinho2024SciRep}, and more recently in quantum metrology~\cite{Porto2024,porto2025NonMarkovian}. Despite their promising features, practical implementations face challenges due to imperfections in generating the desired correlations, arising from limitations in matter-wave sources and focusing techniques~\cite{Zeilinger2002,Viale2003PRA,Marinho_2018}. Recently, a new measure of quantum correlations based on phase-space structure, termed \textit{symplectic coherence}, was introduced to quantify position–momentum correlations in quantum states~\cite{upreti2025symplecticcoherencemeasurepositionmomentum}. Independent of other physical properties of the circuit, recent studies indicate that universal bosonic computations can be classically simulated when the quantum gates employed do not generate position–momentum correlations~\cite{upreti2025interplayresourcesuniversalcontinuousvariable}.

In this work, we investigate the simultaneous estimation of position-momentum (PM) correlations and the effective coupling between a quantum probe and its environment. This approach also allows us to indirectly estimate the temperature, as in a quantum thermometry scheme. Although much of quantum metrology has traditionally focused on pure states and unitary dynamics, fully characterizing realistic systems also requires accounting for decoherence effects. Here, we consider a correlated Gaussian probe undergoing both unitary and non-unitary evolution, and we assess its performance in a multiparameter setting. Specifically, we compare individual versus simultaneous estimation strategies and demonstrate that, even in this more complex framework, the use of initial PM correlations can offer an advantage over uncorrelated Gaussian probes. Our protocol is structured in three stages, using a hybridized parameterization strategy (comprising both unitary and non-unitary evolution): initialization, interaction, and estimation (see Fig. \ref{modelFigure}). Specifically, we estimated the effective temperature of a Markovian bath by examining the environmental coupling using our quantum thermometer, which is based on a single-mode correlated Gaussian system. These correlations influence the elements of the Quantum Fisher Information Matrix (QFIM) after the initial state's evolution through a Markovian bath, and they set the conditions for enhancing our quantum thermometer model with the assistance of PM correlations. Our goal is to compare individual and simultaneous estimation schemes, as well as to establish new bounds in the multi-estimation scenario.

The manuscript is organized as follows. In Section~\ref{sec:theoretical_B}, we review the multiparameter estimation problem, where we present general expressions that are valid for Gaussian systems. Section~\ref{sec:theoretical_C} introduces the correlated Gaussian model for multiparameter estimation. The results are presented in Section~\ref{sec:results}, starting with a comparison between individual and simultaneous estimation strategies (see Section~\ref{sec:results:A}). We then analyze the behavior of the elements of the Quantum Fisher Information Matrix (QFIM) and explore how the precision bounds in the simultaneous estimation scenario can be improved by using a PM-correlated probe as a resource (see Section~\ref{sec:result_B}). Section~\ref{sec:results_C} is dedicated to proving the saturability of the Quantum Cramér–Rao bound for the two parameters under consideration: the PM correlation and the effective temperature. Finally, in Section~\ref{sec:disc}, we discuss the broader implications of our findings.

\section{THEORETICAL FRAMEWORK}\label{sec:theoretical}

This section outlines the theoretical framework for multiparameter estimation in Gaussian quantum systems. Building on this foundation, we introduce a framework for the simultaneous quantum estimation of two key parameters within our model: the position-momentum (PM) correlation and the effective coupling to the environment.

\subsection{Multiparameter Quantum metrology}\label{sec:theoretical_B}

Uncertainty relations impose intrinsic limitations on the precision with which observable quantities can be measured. However, in cases involving parameters not directly associated with observables, quantum estimation theory offers the appropriate framework for analyzing fundamental quantum limits of measurement precision \cite{Monras_IlluminatiPRA2011}. In general, to implement an estimation protocol, one needs to encode a set of parameters $\boldsymbol{\Phi} = \{\phi_1,..., \phi_n \}^{T}$, with $\phi_n$ the $n$th parameter that one wishes to estimate, using the density matrix $ \rho(\boldsymbol{\Phi})$ as a probe. This parameterization process can be implemented through unitary evolution, non-unitary evolution, or a hybrid parameterization strategy~\cite{Liu_JPhysicsA2020}. In multiparameter estimation, the quantum Cramér-Rao bound yields a lower bound to the covariance matrix of the unbiased estimated parameters $\boldsymbol{V}(\boldsymbol{\phi})$ in terms of the quantum Fisher information matrix (QFIM) $\boldsymbol{\mathcal{F}}$ ~\cite{Braunstein_CavesPRL1994,Nichols_AdessoPRA2018,Liu_JPhysicsA2020}
\begin{equation}
\boldsymbol{V}(\boldsymbol{\Phi}) \succeq, \frac{1}{n}\boldsymbol{\mathcal{F}}^{-1},
\label{QCRB}
\end{equation}
where $n$ is the number of repetitions of the experiment. The diagonal elements of $\boldsymbol{V}(\boldsymbol{\Phi})$, i.e., the variances, quantify the error in the estimation of the individual parameters, while the off-diagonal elements indicate the correlations between the parameters. Consequently, QFIM characterizes the fundamental precision limits achievable in multiparameter quantum metrology \cite{Nichols_AdessoPRA2018}. The elements of QFIM are defined as
\begin{equation}
\mathcal{F}_{\phi_i,\phi_j}= \frac{1}{2} \text{Tr} \left[ \rho(\boldsymbol{\Phi}) \{\mathcal{L}_{\phi_i} , \mathcal{L}_{\phi_j}\}  \right],
\end{equation}
where the set of the symmetric logarithmic derivative (SLD) operators $\mathcal{L}_{\phi_n}$ for the parameter $\phi_n$ is implicitly defined
\begin{equation}
    \partial_{\phi_n} \rho (\boldsymbol{\phi}) = \frac{1}{2} \{ \rho (\boldsymbol{\phi}), \mathcal{L}_{\phi_n}  \},
\end{equation}
where $\{\hat{A}, \hat{B}\}$ denotes the anti-commutator of operators $\hat{A}$ and $\hat{B}$. The simplest example of multiparameter estimation involves the simultaneous estimation of two parameters. In this case, $\boldsymbol{\mathcal{F}}^{-1}$ can be calculated explicitly as
\begin{equation}\label{inverseF}
\boldsymbol{\mathcal{F}}^{-1} = \frac{1}{ \boldsymbol{|\mathcal{F}}|} \left ( \begin{array}{cc}
\mathcal{F}_{\phi_j\phi_j} & -\mathcal{F}_{\phi_i\phi_j} \\
-\mathcal{F}_{\phi_i\phi_j} & \mathcal{F}_{\phi_i\phi_i} \\
\end{array} \right),
\end{equation}
where $\boldsymbol{|\mathcal{F}|}= \mathcal{F}_{\phi_i\phi_i}\mathcal{F}_{\phi_j\phi_j}-(\mathcal{F}_{\phi_i\phi_j})^2$. Here, we used the fact that $\boldsymbol{\mathcal{F}}$ is a real symmetric matrix, i.e. $\mathcal{F}_{\phi_i\phi_j}=\mathcal{F}_{\phi_j\phi_i}$. The main diagonal elements $\mathcal{F}_{\phi_i,\phi_i}$ correspond to the well-known quantum Fisher information (QFI), describing how much information is available to estimate each parameter independently. However, to quantify the precision limits in a simultaneous estimation scenario, we must account for the off-diagonal elements $\mathcal{F}_{\phi_i\phi_j}$, which reflect parameter correlations. By combining the Quantum Cramér-Rao bound in Eq.~(\ref{QCRB}) with the explicit inverse matrix in Eq.~(\ref{inverseF}), we can define the \textit{effective} Fisher information terms $\widetilde{\mathcal{F}}$ for each element of the covariance matrix. These quantities, which incorporate the correlation terms, are defined implicitly by $(\boldsymbol{V})_{ij} \geq (n\widetilde{\mathcal{F}}_{ij})^{-1}$, yielding the explicit forms:
\begin{gather}
\widetilde{\mathcal{F}}_{\phi_i\phi_i}  = \mathcal{F}_{\phi_i\phi_i} - \mathcal{F}_{\phi_i\phi_j}^2/ \mathcal{F}_{\phi_j\phi_j} \;, \qquad \widetilde{\mathcal{F}}_{\phi_i\phi_j} =\mathcal{F}_{\phi_i\phi_j} - \mathcal{F}_{\phi_i\phi_i}\mathcal{F}_{\phi_j\phi_j}/ \mathcal{F}_{\phi_i\phi_j}, \nonumber\\
 \qquad  \widetilde{\mathcal{F}}_{\phi_i\phi_j}= \mathcal{F}_{\phi_i\phi_j} - \mathcal{F}_{\phi_i\phi_i}\mathcal{F}_{\phi_j\phi_j}/ \mathcal{F}_{\phi_i\phi_j}, \qquad  \widetilde{\mathcal{F}}_{\phi_j\phi_j} = \mathcal{F}_{\phi_j\phi_j} - \mathcal{F}_{\phi_i\phi_j}^2/ \mathcal{F}_{\phi_i\phi_i}. \label{eq:new_bounds}
\end{gather}
Note that, unlike single-parameter estimation, where Quantum Fisher Information (QFI) sets the ultimate precision bound, the multiparameter case involves trade-offs: the off-diagonal elements of the Quantum Fisher Information Matrix (QFIM) reflect parameter correlations that can reduce the precision achievable for each individual parameter, due to the simultaneous nature of their estimation~\cite{Szczykulska2016Review}. Under these conditions, the accuracy of estimating one parameter is affected by the presence of the other unknown parameter. Moreover, to estimate each parameter with the same level of precision as if the remaining parameters were known exactly, the parameters must be statistically independent, i.e. $\mathcal{F}_{\phi_i\phi_j}=0$ for $i\neq j$, indicating that the Fisher information matrix is diagonal. This condition guarantees that the uncertainty associated with one parameter does not influence the precision with which the others can be estimated~\cite{Szczykulska2016Review}.

Restricting the analysis to Gaussian states~\cite{CavesPRL2013,PinelPRA2013,monras2013ARXIV,Safranek_2015NJP,Safranek_2016PRA,Jonas_2024,Jonas_2025,Pritam2025generictwomodegaussianstates}, which are fully characterized by their first moments, given by the displacement vector $\boldsymbol{d}=\langle \boldsymbol{r} \rangle$, where $\boldsymbol{r}$ denotes the vector of quadrature operators with componets $\hat{r}_i$, and their second moments, represented by the covariance matrix $\boldsymbol{\sigma}_{ij}=\langle \{ \Delta\hat{r}_i, \Delta\hat{r}_j  \}\rangle$, with $ \Delta\hat{r}_i = \hat{r}_i -\langle \hat{r}_i\rangle$, we can compute the elements of the quantum Fisher information matrix (QFIM) in terms of these first and second moments~\cite{Gao2014EPJD,Safranek_2019JPA_Math_Theor}
\begin{gather}\label{eq:Dominik}
\mathcal{F}_{\phi_i\phi_j}=\frac{1}{2} \textbf{\text{vec}}[\partial_{\phi_i}\boldsymbol{\sigma}]^{\text{T}} \boldsymbol{\mathcal{M}}^{-1}\textbf{\text{vec}}[\partial_{\phi_j}\boldsymbol{\sigma}]+2(\partial_{\phi_i}\boldsymbol{d})^{\text{T}}(\boldsymbol{\sigma^{-1}})(\partial_{\phi_j}\boldsymbol{d}),
\end{gather}
where $\boldsymbol{\mathcal{M}}= \boldsymbol{\sigma} \otimes \boldsymbol{\sigma} - \boldsymbol{\Omega} \otimes \boldsymbol{\Omega}$, with $\boldsymbol{\Omega}$ being the symplectic matrix defined by the commutation relations $[\hat{r}_k, \hat{r}_l] = i \Omega_{kl}$~\cite{serafini2017quantum} (for a single-mode system, $\boldsymbol{\Omega} = i\sigma_y$, with $\sigma_y$ being the Pauli-$y$ matrix). Here, $\otimes$ denoting the Kronecker product, $\partial_{\phi_i}$ represents the derivative with respect to the parameter $\phi_i$, and the operation $\textbf{\text{vec}}[\cdot]$ denotes the vectorization of a matrix~\cite{Safranek_2019JPA_Math_Theor}. The first term is associated with the dynamical dependence of the covariance matrix on the parameters $\phi_i$ and $\phi_j$, and the second is the contribution of the displacement-vector dynamics of the Gaussian state to the estimated parameters. 

Following previous approaches~\cite{Yousefjani_AdessoPRA2017,Nichols_AdessoPRA2018}, we employed a ratio to quantitatively compare the metrological performance of individual and simultaneous estimation schemes  
\begin{equation}
\mathcal{R} = \frac{\Delta_I}{\Delta_S},
\end{equation}
where (considering a single repetition, $n=1$)
\begin{equation}\label{eq:total_variances}
    \Delta_I = \sum_{\phi_i \in \boldsymbol{\phi}} \mathcal{F}_{\phi_i\phi_i}^{-1} \;\;\;\;\;\;\; \text{and} \;\;\;\;\; \Delta_S = \kappa^{-1} \text{Tr} [\mathcal{F}^{-1} ],
\end{equation}
represents the minimal total variances in the individual and simultaneous cases, respectively. Within this definition, $\kappa$ represents the total number of estimated parameters. Therefore, the $\kappa^{-1}$ factor is included to account for the fact that the simultaneous scheme requires $\kappa$ fewer resources than individually estimating each parameter while resetting the probe each time. This ratio can thus be regarded as a figure of merit for assessing the performance of individual and simultaneous strategies. Simultaneous estimation is more advantageous than individual estimation when $\mathcal{R}>1$, whereas individual estimation is preferable when $\mathcal{R}<1$. The maximum advantage of simultaneous estimation, $\mathcal{R}_{max}=\kappa$, is achievable only when the condition $\mathcal{F}_{\phi_i\phi_j}=0$ for $i\neq j$ holds~\cite{Yousefjani_AdessoPRA2017,Nichols_AdessoPRA2018,Braun_PirandolaRevModPhys2018}.

While the Cramér-Rao bound is asymptotically attainable in individual parameter estimation, the simultaneous estimation of multiple parameters does not necessarily offer an advantage over estimating each parameter individually \cite{RagyPRA2016, Nichols_AdessoPRA2018}. This is because achieving the Cramér-Rao bound in the multiparameter case requires an optimal measurement strategy. A sufficient criterion is met when the optimal measurement for each parameter corresponds to a set of projectors that commute with the respective symmetric logarithmic derivative (SLD), that is, when $[\mathcal{L}_{\phi_i}, \mathcal{L}_{\phi_j}] = 0$ \cite{Nichols_AdessoPRA2018}.  This commutation relation ensures the existence of a common eigenbasis for the SLDs, enabling the construction of a measurement that is optimal for estimating both parameters $\phi_i$ and $\phi_j$ simultaneously. A less stringent condition asserts that the multiparameter Cramér-Rao bound can be asymptotically saturated if all pairs of SLDs commute on average, that is, when their expectation value concerning the quantum state vanishes \cite{Nichols_AdessoPRA2018,RagyPRA2016} 
\begin{equation}
   \mathrm{Tr}(\rho [\mathcal{L}_{\phi_i}, \mathcal{L}_{\phi_j}]) = 0.
\end{equation}
This requirement defines the optimal measurement compatibility condition (ii) between a pair of parameters \cite{RagyPRA2016}. In the context of Gaussian states, this condition can be explicitly expressed as \cite{Safranek_2019JPA_Math_Theor}
\begin{equation}\label{eq:Safranek_2019JPA_Math_Theor}
   \mathrm{Tr}(\rho [\mathcal{L}_{\phi_i}, \mathcal{L}_{\phi_j}]) = \text{vec}[\partial_{\phi_i}\boldsymbol{\sigma}]^{\text{T}} \mathcal{M}^{-1}(\boldsymbol{\sigma}\otimes \boldsymbol{\Omega}-\boldsymbol{\Omega}\otimes \boldsymbol{\sigma})\mathcal{M}^{-1}\text{vec}[\partial_{\phi_j}\boldsymbol{\sigma}]+4(\partial_{\phi_i}\boldsymbol{d})^{\text{T}}(\boldsymbol{\sigma^{-1}}\boldsymbol{\Omega} \boldsymbol{\sigma^{-1}})(\partial_{\phi_j}\boldsymbol{d}).
\end{equation}
Again, since for a Gaussian state the system is fully characterized by its first moments (the displacement vector $\boldsymbol{d}$) and second moments (the covariance matrix $\boldsymbol{\sigma}$), the condition (\ref{eq:Safranek_2019JPA_Math_Theor}) is determined solely by these statistical quantities.

\subsection{Estimation protocol}\label{sec:theoretical_C}

\begin{figure}[ht]
\centering
\includegraphics[scale = 1.0]{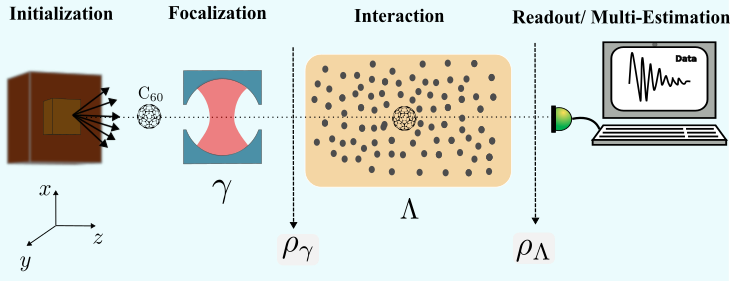}
\caption{Protocol to describe the simultaneous estimation of the parameters $\Lambda$ (effective environmental coupling constant) and $\gamma$ (position–momentum correlations). Fullerene wave packets with momentum $\vec{k}$ and an initial coherent length $\ell_0$ are produced during the initialization procedure. The focalization stage then takes place, producing initial position–momentum correlations, represented by $\gamma$. After then, there is interaction with a Markovian thermal reservoir, and the effective environmental coupling is quantified by $\Lambda$. Multiple parameters are estimated simultaneously during the readout process.}
\label{modelFigure}
\end{figure}

The three main stages of the quantum estimation protocol are initialization, interaction, and readout-estimation. These steps are schematically shown in Fig.~\ref{modelFigure}. Fullerene Gaussian wave packets with momentum $\vec{k}$ and initial coherence length $\ell_0$ are prepared during the initialization process. Two procedures comprise the parameterization. Focalization is a unitary process that uses a real-noisy source to encode the position-momentum (PM) correlation parameter $\gamma$ (for more information on how PM can be generated, see Ref.~\cite{Porto2024}). Interaction is a non-unitary process in which the probe couples to a Markovian bath, encoding the effective coupling constant $\Lambda$ and, consequently, its temperature $T$. In order to simultaneously infer the unknown parameters, including (i) PM correlations and (ii) thermometry, the estimate step is finally performed. This includes traditional post-processing of the measurement results. 

We consider a position-momentum (PM) correlated Gaussian state produced by a non-ideal source with a transverse width $\sigma_{0}$, which corresponds to the focalization stage depicted in Fig.~\ref{modelFigure} as the initial state~\cite{dodonov2002nonclassical}.
\begin{equation}\label{psi_0}
\rho_{\gamma}(x_{0},x_{0}^{\prime}) = \frac{1}{\sqrt{\pi}\sigma_{0}} \exp\left[-(1-i\gamma)\frac{x_{0}^{2}}{2\sigma_{0}^{2}}\right] \exp\left[-(1+i\gamma)\frac{x_{0}^{\prime 2}}{2\sigma_{0}^{2}}\right] \exp\left[-\frac{(x_{0}-x_{0}^{\prime})^{2}}{2\ell_{0}^{2}}\right]. 
\end{equation}
The spatial coherence of the generated quantum state is measured by the parameter $\ell_{0}$, which describes the degree of beam collimation along the transverse axis $Ox$. The limit $\ell_{0} \to \infty$ represents perfect collimation, which corresponds to a fully coherent state. On the other hand, the limit $\ell_{0} \to 0$~\cite{PP2024} characterizes the state that results from the lack of collimation, which is completely incoherent. $\gamma$ is the real parameter that controls the existence of correlations in the initial state ~\cite{dodonov2002nonclassical,dodonov2014transmission}. The position-momentum correlation for the state specified in Eq.(\ref{psi_0}) is $\sigma_{xp}= \langle ( \hat{x}\hat{p} +\hat{p}\hat{x}) \rangle -2\langle \hat{x} \rangle \langle \hat{p} \rangle =\hbar\gamma$. The system represents an uncorrelated Gaussian wave packet when $\gamma = 0$, which means that $\sigma_{xp} = 0$. In this case, the initial state reduces to a standard uncorrelated coherent Gaussian state with uncertainties that are the same as the vacuum state~\cite{Marinho2024SciRep,porto2025NonMarkovian}. For a detailed analysis connecting the correlation parameter and the squeezing of the phase-space distributions, see Refs.~\cite{porto2025NonMarkovian,ThiagoNJP_2025,dearaujo2025gravimetryenhancementsqueezedstates}.

In this work, we explore how initial position-momentum (PM) correlations could be employed as a quantum resource to improve multiparameter estimation in noisy settings. Since environmental scattering decoherence, caused by ambient particles and fields, including air molecules and cosmic background radiation $3$ K~\cite{Schlosshauer2}, continuously monitors quantum positions, we include this effect in the interaction step of the protocol (Fig.~\ref{modelFigure}). After the focalization process has ended, the system interacts with a Markovian bath~\cite{PP2024}, and its evolution is governed by Feynman's propagator that includes the environmental effect ~\cite{Viale2003PRA,Schlosshauer2}
\begin{gather}
G_{\Lambda}(x,x^{\prime},t;x_0,x^{\prime}_0,0)=\frac{m}{2\pi\hbar t}\;\exp{\Big\{\frac{im}{2\hbar t}\left[(x-x_0)^{2}+(x^{\prime}-x_0)^{2}\right]-\frac{\Lambda t}{3}\left[(x-x^{\prime})^{2}+(x_0-x^{\prime}_{0})^{2}+(x-x^{\prime})(x_0-x^{\prime}_0)\right]\Big\}}.\label{propagator}
\end{gather}
therefore, the density matrix evolution is given by
\begin{equation}
\rho_{\Lambda}(x,x^{\prime},t)=\int \int dx_0 dx^{\prime}_0 G_{\Lambda}(x,x^{\prime},t;x_0,x^{\prime}_0,0)\rho_{\gamma}(x_0,x^{\prime}_0).
\label{rho_x}
\end{equation}
The parameter $\Lambda$ characterizes how quickly spatial coherence over a distance $\Delta x$ is lost, leading to a decoherence timescale defined as $\tau_{\text{dec}}^{-1} = \Lambda (\Delta x)^2$~\cite{Schlosshauer2}. Considering the decoherence effect caused by the scattering of air molecules and the propagation of fullerene molecules~\cite{Viale2003PRA}, the effective scattering constant in the long-wavelength limit is provided by $\Lambda(T)= (8/3\hbar^2)\sqrt{2\pi m_{\text{air}}}(k_B T)^{3/2}N w^2$ \cite{Schlosshauer2}, where $N$ is the total number density of the air, $m_{\text{air}}$ is the mass of the air molecule, $w$ is the size of the molecule of the quantum system, $k_B$ is the Boltzmann constant, and $T$ is the bath temperature~\cite{Schlosshauer2}. This decoherence model, arising from air molecule scattering, remains a subject of active investigation. In a recent study, the effective scattering rate constant $\Lambda$ was re-evaluated using a series expansion, avoiding approximations related to the wavelength of air molecules~\cite{MazumdarPRA2025}. After integrating and manipulating Eq.~(\ref{rho_x}) algebraically, we get
\begin{equation}\label{rho_Lambda}
\rho_{\Lambda}(x,x^{\prime},t)=\mathcal{N}_{t}\exp\left\{ -\mathcal{A}_{t}x^{2}-\mathcal{B}_{t}x^{\prime2}+\mathcal{C}_{t}xx^{\prime}\right\}, 
\end{equation}
where $\mathcal{A}_{t}$, $\mathcal{B}_{t}$, $\mathcal{C}_{t}$ and $\mathcal{N}_{t}$ are parameters that include the interaction with the Markovian bath (see Appendix \ref{appendix:rho_parameters}).

The dimensionless displacement vector $\boldsymbol{d} = (\langle \hat{x} \rangle, \langle \hat{p} \rangle)$ vanishes for the mixed Gaussian state defined in Eq.~(\ref{rho_Lambda}), and the corresponding covariance matrix $\boldsymbol{\sigma}$ is given by
\begin{equation}\label{eq:Cov_matrix}
\boldsymbol{\sigma}=\left(\begin{array}{cc}
\sigma_{xx} & \sigma_{xp}\\
\sigma_{px} & \sigma_{pp}
\end{array}\right),
\end{equation}
with
\begin{equation}\label{eq:sigmaxx_sigampp}
\sigma_{xx}=\frac{\langle \hat{x}^2\rangle - \langle \hat{x}\rangle^2}{\sigma_0^2}= \frac{t^{2}}{\tau_{0}^{2}}\left[1+\left(\frac{\tau_{0}}{t}+\gamma\right)^{2}+\frac{2\sigma_{0}^{2}}{\ell_{0}^{2}}+\frac{4\Lambda t\sigma_{0}^{2}}{3}\right], \;\;\;\;\;\;  \sigma_{pp} =\frac{\langle \hat{p}^2\rangle - \langle \hat{p}\rangle^2}{(\hbar^2/\sigma_0^2)}=1+\gamma^{2}+\frac{2\sigma_{0}^{2}}{\ell_{0}^{2}}+4t\Lambda\sigma_{0}^{2}, 
\end{equation}
\begin{gather}
     \sigma_{px}= \sigma_{xp}=\frac{\langle ( \hat{x}\hat{p} +\hat{p}\hat{x}) \rangle -2\langle \hat{x} \rangle \langle \hat{p} \rangle}{\hbar}=\gamma+\frac{t}{\tau_{0}}(1+\gamma^{2})+\frac{2t\sigma_{0}^{2}}{\tau_{0}\ell_{0}^{2}}+\frac{2t^{2}\Lambda\sigma_{0}^{2}}{\tau_{0}}, \label{eq:sigmaxp}
\end{gather}
where $\tau_0 = (m\sigma_0^2)/\hbar$ is a constant with time dimension \cite{Marinho_2020}. 

To enhance the clarity and interpretation of the results, specifically, how position-momentum correlations enhance the Fisher information, we introduce the Wigner function as a phase-space representation of the quantum state \cite{Wigner1932}. This function offers a quasiprobability distribution in the classical phase space $(x, p)$. It is widely recognized as an effective tool for visualizing quantum states \cite{BanaszekPRA1998}, and it is defined as~\cite{leonhardt1997measuring}
\begin{equation}
W(x,p) = \frac{1}{\pi\hbar}\int_{-\infty}^{\infty} dy \; e^{2ipy/\hbar} \rho (x-y, x+y),
\end{equation}
providing a full representation of the state in phase space. In the case of Gaussian states, the Wigner function retains a Gaussian form and takes the explicit expression \cite{RevModPhys2012Lloyd}
\begin{equation}
 W (\boldsymbol{r}) =\frac{ \exp\left[-\frac{1}{2}(\boldsymbol{r}-\boldsymbol{d})^{\text{T}}\boldsymbol{\sigma}^{-1}(\boldsymbol{r}-\boldsymbol{d})\right]}{(2\pi) \sqrt{\text{det}\boldsymbol{\sigma}}},
\end{equation}
where $\boldsymbol{r} = (x, p)$ represents the phase-space coordinate vector. As such, the quantum state is fully characterized by its first and second statistical moments: the displacement vector $\boldsymbol{d}$ and the covariance matrix $\boldsymbol{\sigma}$ (\ref{eq:Cov_matrix}), respectively.

\section{Results}\label{sec:results}

We begin analyzing the behavior of the QFIM throughout the different stages of the parameterization process, as illustrated in Fig.~\ref{modelFigure}, and the performance of individual and simultaneous estimation in Sec.~\ref{sec:results:A}. Then, Sec.~\ref{sec:result_B} focuses on the estimation of position–momentum (PM) correlations in the presence of unavoidable environmental interactions, and examines the impact of simultaneously estimating the effective scattering constant, $\Lambda(T)$. Also, we investigate how these initial correlations can enhance precision in noisy multiparameter quantum metrology. Finally, Sec.~\ref{sec:results_C} analyzes the saturability of the Quantum Cramér–Rao bound for the PM correlation and effective temperature.

\subsection{Individual \textit{versus} Simultaneous estimation}\label{sec:results:A}

In the subsequent analysis, we calculate the elements of the Quantum Fisher Information Matrix (QFIM), as given by Eq. (\ref{eq:Dominik}) (see Appendix \ref{appendix:QFIM} for explicit expressions), setting $\phi_i=\gamma$ and $\phi_j=\Lambda(T)$. Consequently, the minimal total variances (\ref{eq:total_variances}) in the individual and simultaneous estimation scenarios can be expressed as
\begin{gather}
\Delta_I = \mathcal{F}_{\gamma\gamma}^{-1}+\mathcal{F}_{\Lambda\Lambda}^{-1}, \;\;\; \text{and} \;\;\;\;\; \Delta_S = 2^{-1}( \widetilde{\mathcal{F}}_{\gamma\gamma}^{-1} +\widetilde{\mathcal{F}}_{\Lambda\Lambda}^{-1}),
\end{gather}
with $\kappa=2$ in our case. We can finally evaluate the performance ratio $\mathcal{R}=\Delta_I/\Delta_S$ from these two quantities.

\begin{figure*}[!hbt]
\centering
\includegraphics[scale=0.44]{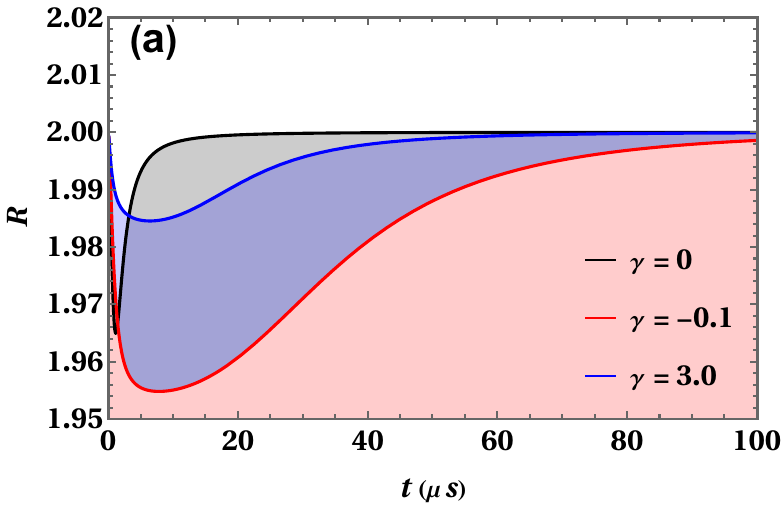}
\includegraphics[scale=0.44]{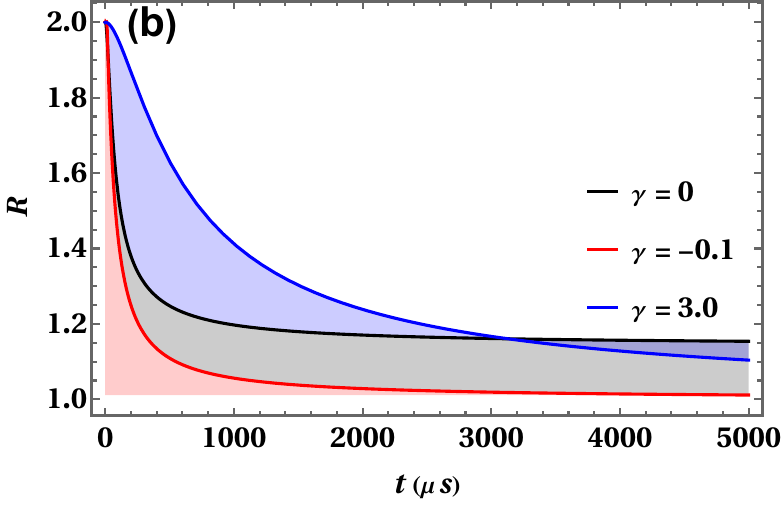}
\includegraphics[scale=0.44]{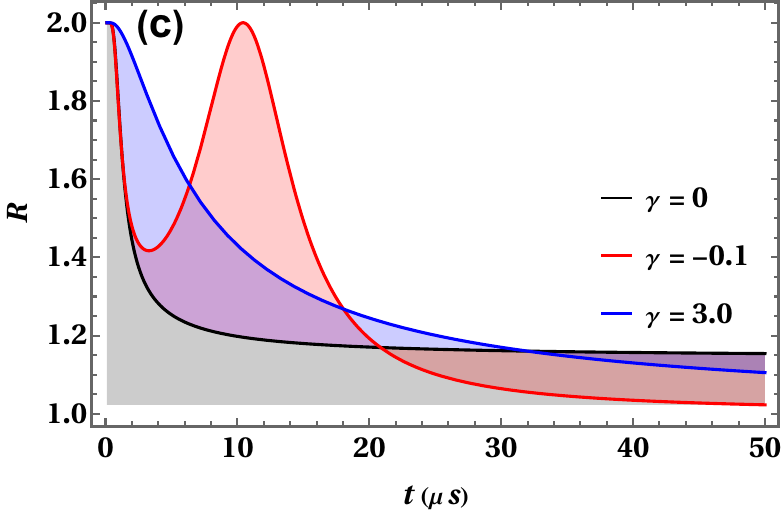}
\caption{Temporal behavior of the performance ratio $\mathcal{R}$ for increasing values of the environmental effective coupling constant (effective temperature): (a) $\Lambda = 3 \times 10^{15}$ m$^{-2}$s$^{-1}$ ($T=16.9$ mK), (b) $\Lambda = 3 \times 10^{20}$ m$^{-2}$s$^{-1}$ ($T=36.5$ K), and (c) $\Lambda = 3 \times 10^{22}$ m$^{-2}$s$^{-1}$ ($T=786$ K), demonstrating the mitigation of decoherence in simultaneous estimation via initial correlations $\gamma$ within specific time windows.}\label{Fig2}
\end{figure*}
All subsequent plots employ the following parameters~\cite{Brezger_PhysRevLett2022,Hornberger_PhysRevLett2003}: mass of fullerene $m=1.2\times 10^{-24}$ kg, molecular size $w = 7$ $\mathring{A}$, initial wave packet width $\sigma_0 = 7.8$ nm, mass of the air molecule $m_{\text{air}}=5.0 \times 10^{-26}$ kg, and density of the air molecules $N = 4.0\times 10^{14}$ molecules/m$^3$~\cite{Marinho_2018,Marinho_2023,Viale2003PRA}, and initial coherence length $\ell_{0} = 50\;\mathrm{nm}$. 
\begin{figure*}[!hbt]
\centering
\includegraphics[scale=0.35]{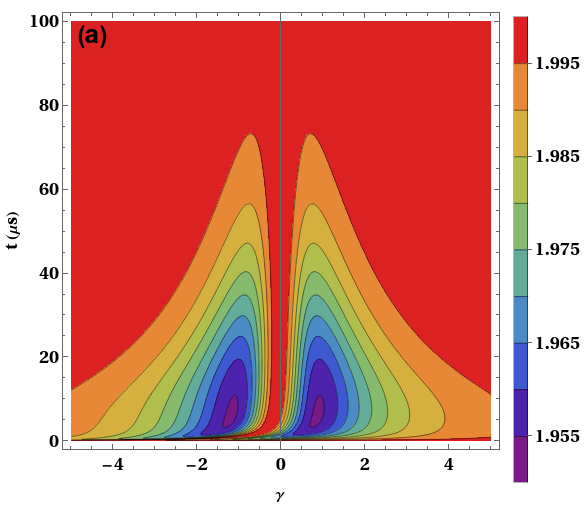}
\includegraphics[scale=0.35]{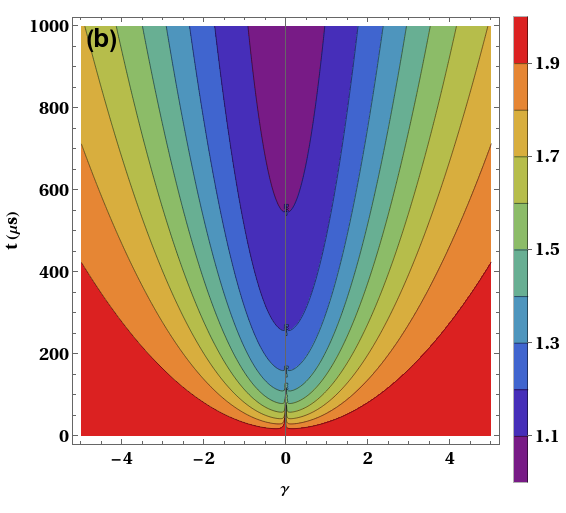}
\includegraphics[scale=0.35]{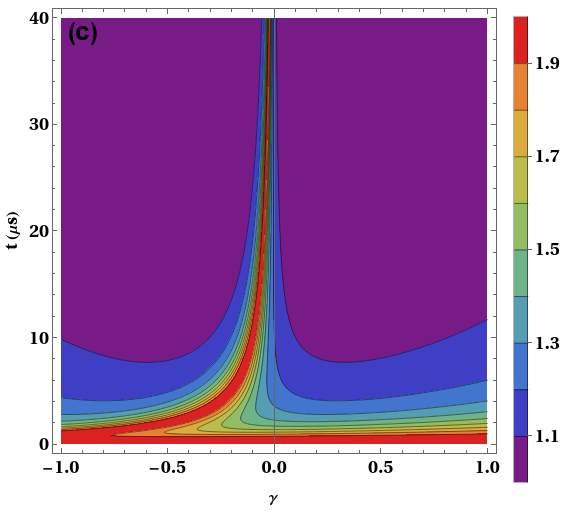}
\caption{Performance ratio $\mathcal{R}$ as a function of time and initial correlation $\gamma$ for increasing values of the environmental effective constant: (a) $\Lambda = 3 \times 10^{15}$ m$^{-2}$s$^{-1}$ ($T=16.9$ mK), (b) $\Lambda = 3 \times 10^{20}$ m$^{-2}$s$^{-1}$ ($T=36.5$ K), and (c) $\Lambda = 3 \times 10^{22}$ m$^{-2}$s$^{-1}$ ($T=786$ K), identifying a clear parameter window where simultaneous estimation outperforms the individual strategy, even under conditions of strong decoherence.}\label{Fig3}
\end{figure*}
Figure~\ref{Fig2} illustrates the temporal behavior of the performance ratio $\mathcal{R}$ with increasing values of the effective environmental coupling constant $\Lambda(T)$. Specifically, three scenarios are considered: (a) a weak environmental influence with $\Lambda = 3 \times 10^{15}\mathrm{m}^{-2}\mathrm{s}^{-1}$, (b) an intermediate regime with $\Lambda = 3 \times 10^{20}\mathrm{m}^{-2}\mathrm{s}^{-1}$, and (c) a strong environmental influence characterized by $\Lambda = 3 \times 10^{22}\mathrm{m}^{-2}\mathrm{s}^{-1}$. In each case, the performance ratio $\mathcal{R}$ is evaluated as a function of time for different values of the initial correlation parameter $\gamma$. The results demonstrate how increasing environmental coupling (effective temperature) impacts the temporal evolution of the performance ratio in simultaneous parameter estimation, highlighting the role of initial correlations in mitigating the detrimental effects of decoherence within a specific temporal range. The contour plots of the performance ratio $\mathcal{R}$ as a function of time and the initial correlation parameter $\gamma$, shown in Fig.~\ref{Fig3}, support this observation by revealing a region where simultaneous estimation remains effective even under strong decoherence, provided that an appropriate parameter regime is selected.

\subsection{QFIM elements behavior}\label{sec:result_B}

\begin{figure*}[!htb]
\centering
\includegraphics[scale = 0.35]{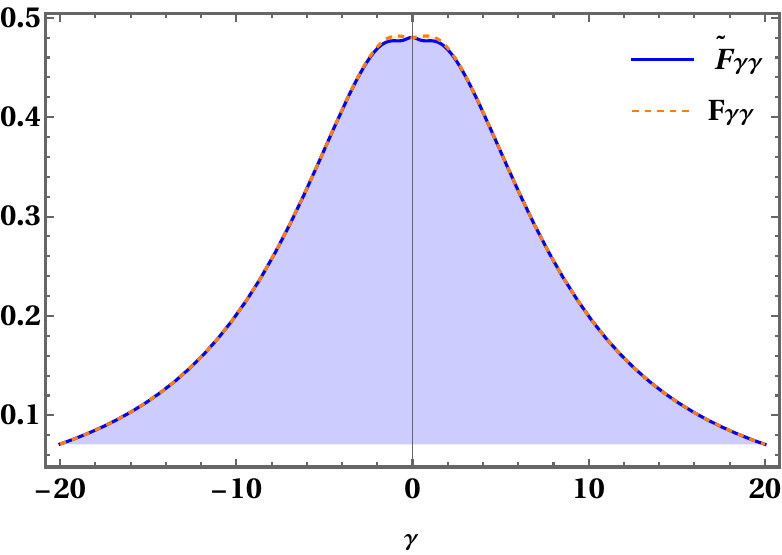}
\includegraphics[scale = 0.35]{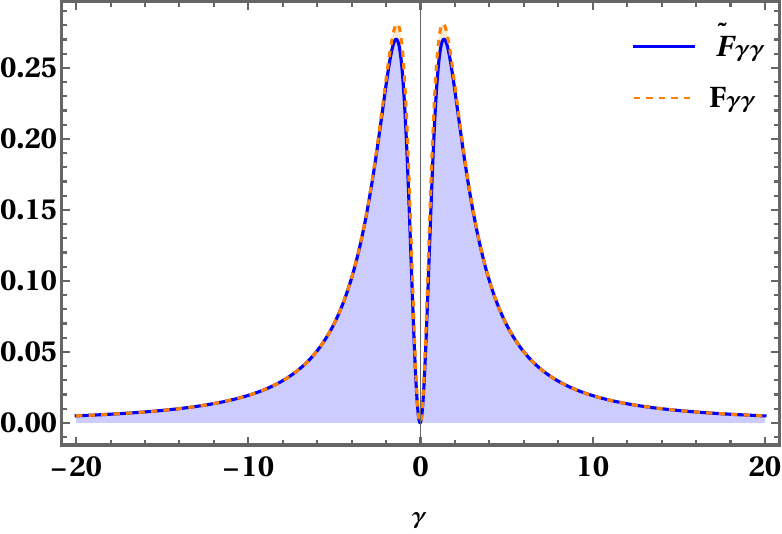}
\includegraphics[scale = 0.35]{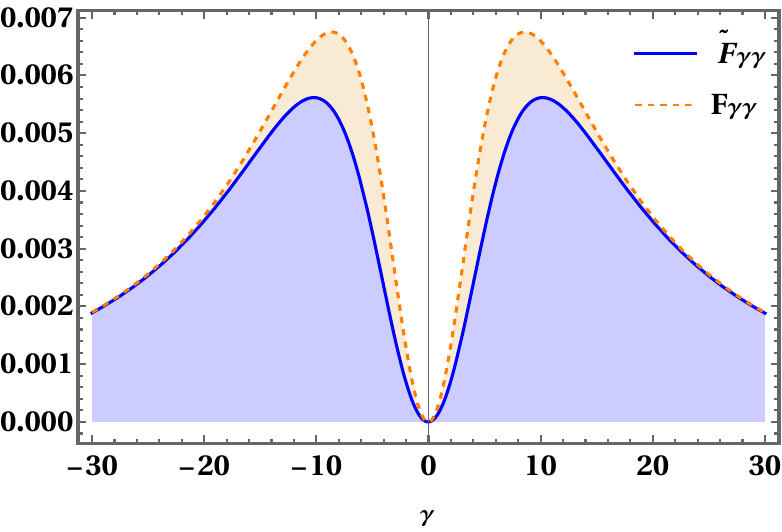}\\
\includegraphics[scale = 0.35]{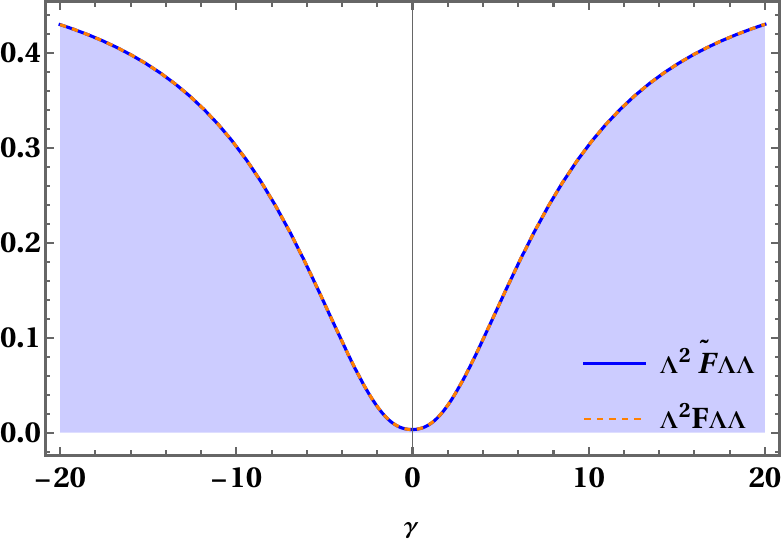}
\includegraphics[scale = 0.35]{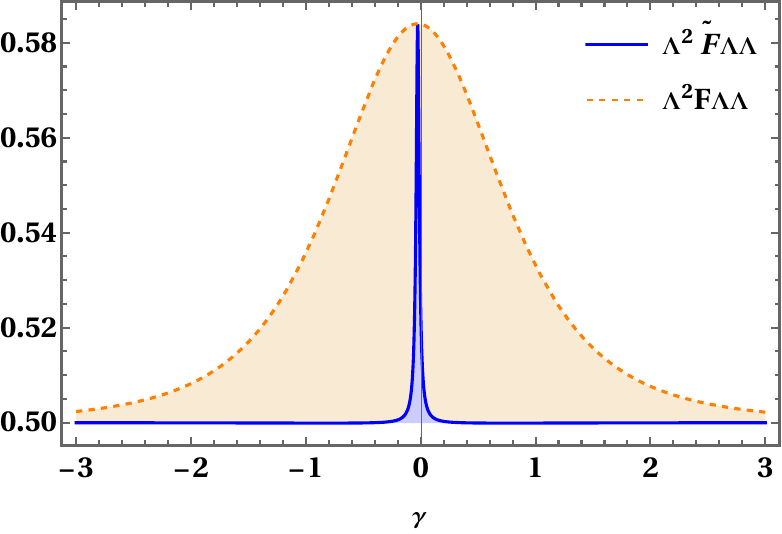}
\includegraphics[scale = 0.35]{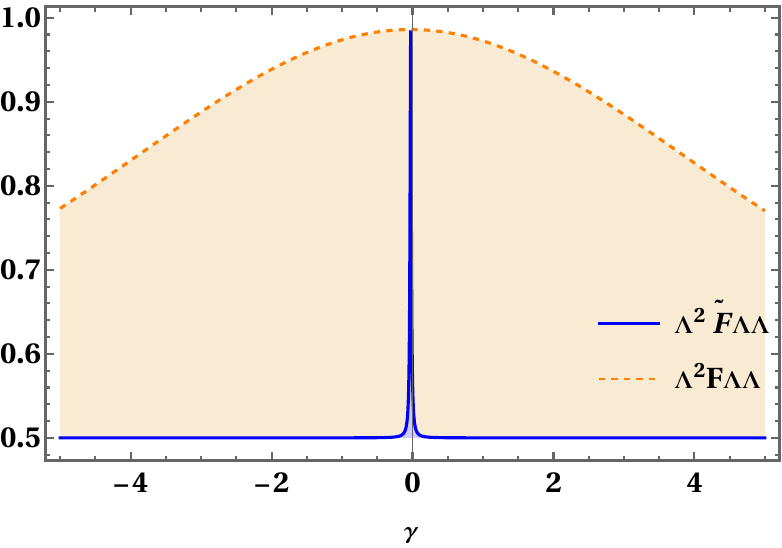}\\
\includegraphics[scale = 0.35]{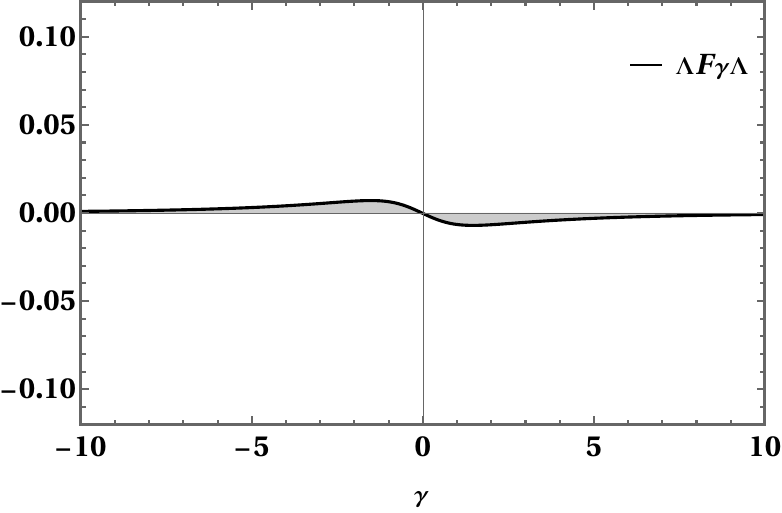}
\includegraphics[scale = 0.35]{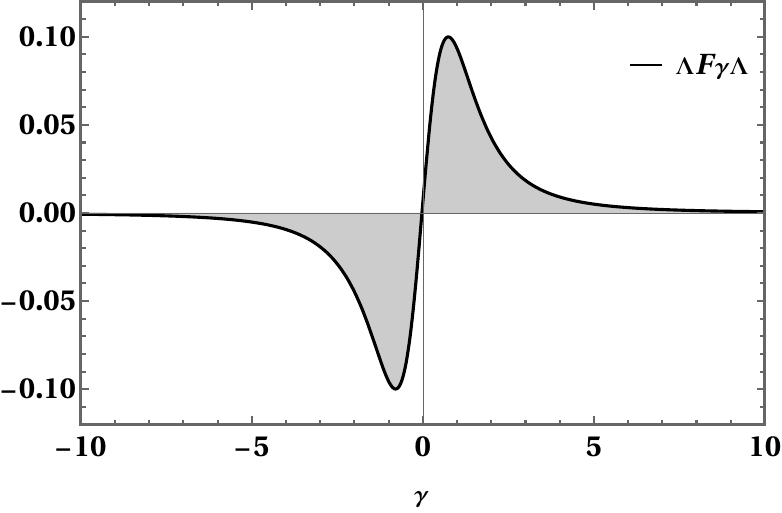}
\includegraphics[scale = 0.35]{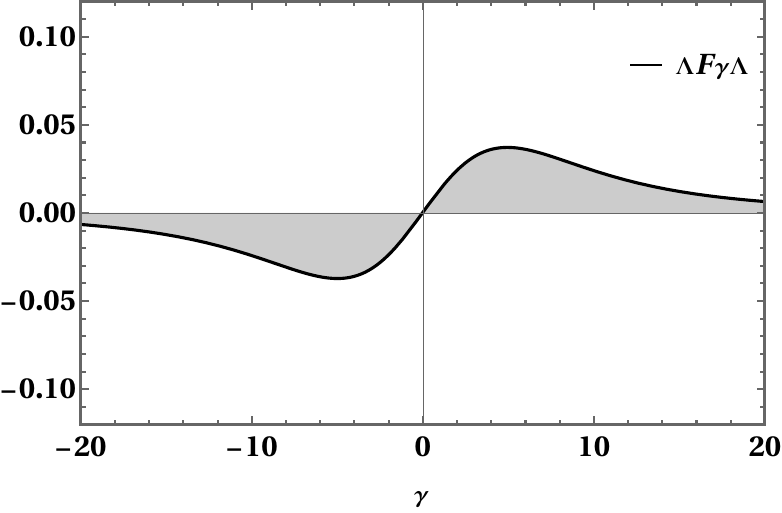}
\caption{The behavior of the Quantum Fisher Information Matrix (QFIM) elements is shown. The solid lines represent the newly derived bounds, denoted as $\widetilde{\mathcal{F}}_{ij}$, corresponding to the framework of simultaneous multiparameter estimation. In contrast, the dashed lines indicate the bounds $\mathcal{F}_{ij}$ associated with the individual estimation of each parameter. The plots from left to right correspond to three distinct regimes: a weak environmental effect regime with $\Lambda = 3 \times 10^{15}\mathrm{m}^{-2}\mathrm{s}^{-1}$ ($T=16.9$ mK - left), an intermediate regime with $\Lambda = 3 \times 10^{20}\mathrm{m}^{-2}\mathrm{s}^{-1}$ ($T=36.5$ K - center), and a strong environmental effect regime with $\Lambda = 3 \times 10^{22}\mathrm{m}^{-2}\mathrm{s}^{-1}$ ($T=786$ K - right). In all cases, the interaction time is fixed at $t = 40\mu\mathrm{s}$. Our analysis of the newly derived bounds confirms that, given appropriate initial correlations $\gamma$, the simultaneous precision can attain the optimal individual bound.}\label{Fig4}
\end{figure*}

\begin{figure*}[!ht]
\centering
\centering
\includegraphics[scale=0.4]{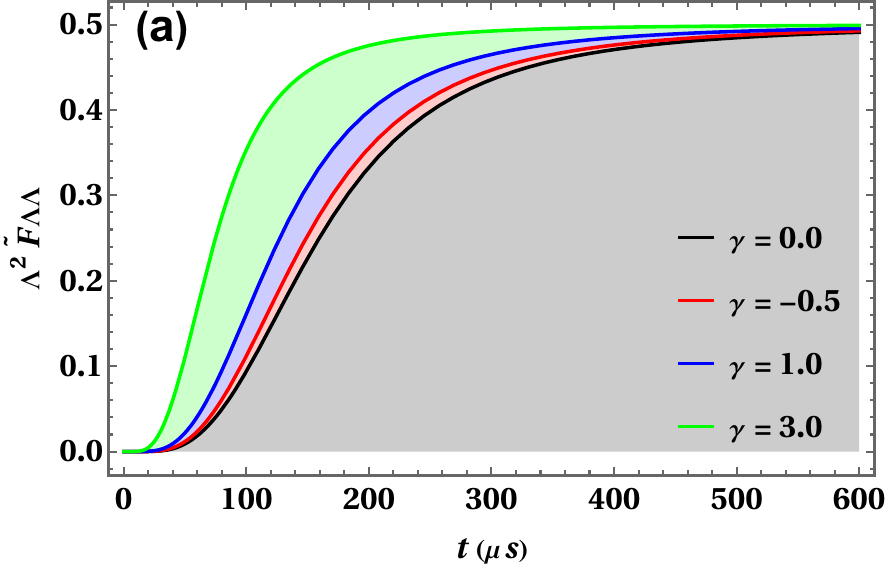}
\includegraphics[scale=0.4]{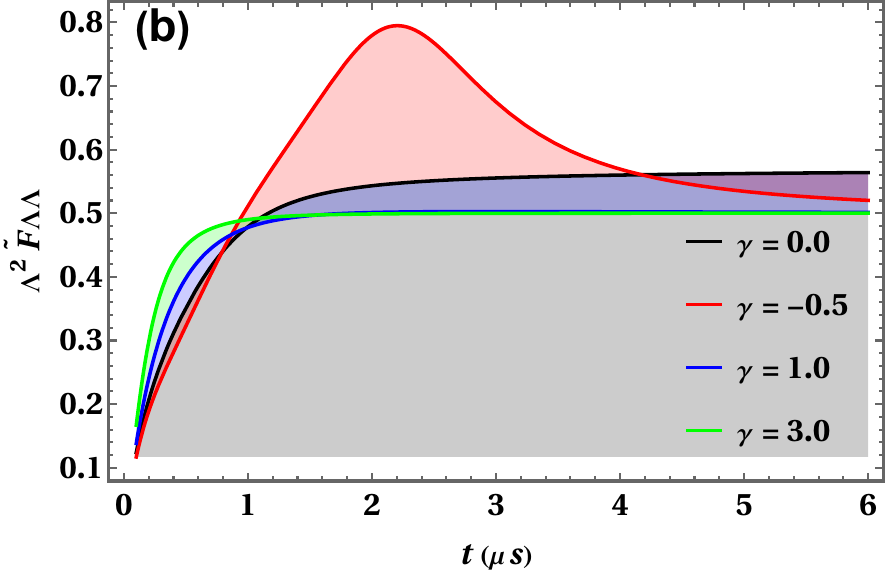}
\caption{Temporal behavior of the newly derived bound $\widetilde{\mathcal{F}}_{\Lambda \Lambda}$ for effective temperature estimation is analyzed within the framework of simultaneous multiparameter estimation for different values of the initial correlation parameter $\gamma$. Panels (a) correspond to a weak environmental effect, characterized by $\Lambda = 3 \times 10^{15}\mathrm{m}^{-2}\mathrm{s}^{-1}$ ($T=16.9$ mK), while (b) represent a strong environmental effect with $\Lambda = 3 \times 10^{22}\mathrm{m}^{-2}\mathrm{s}^{-1}$ ($T=786$ K). It is important to emphasize that, in the regime of effective weak coupling (a), the newly derived precision bounds are practically indistinguishable from those obtained using the individual estimation scheme analyzed in Ref.~\cite{Porto2024}, with both curves being superposed and, for this reason, only one is shown in the figure.}\label{Fig5}
\end{figure*}

Now we examine the dynamics of each element of the QFIM (see Appendix \ref{appendix:QFIM} for explicit expressions). Specifically, we compare the evolution of the QFIM elements $\mathcal{F}_{\gamma\gamma}$ and $\mathcal{F}_{\Lambda\Lambda}$ under individual estimations, alongside the modified bounds $\widetilde{\mathcal{F}}_{\gamma\gamma}$ and $\widetilde{\mathcal{F}}_{\Lambda\Lambda}$ obtained in the case of simultaneous estimations. Figure~\ref{Fig4} illustrates the behavior of the elements of the QFIM,  highlighting both the newly derived bounds $\widetilde{\mathcal{F}}_{ij}$~(solid-lines), relevant to simultaneous multiparameter estimation, and the standard bounds $\mathcal{F}_{ij}$~(dotted-lines), which correspond to the individual estimation of each parameter. The analysis of the curves provides evidence that in most cases, the precision achieved by multiparameter estimation is consistently lower than that achieved by single-parameter estimation. These differences indicate that the precision bound associated with the estimation of a single-parameter provides an ultimate upper bound for the more general multiparameter scenario, as can be seen directly from the equation~(\ref{eq:new_bounds}), where the off-diagonal elements of QFIM reduce the information available for estimating each parameter due to the simultaneous measurement and determination of these parameters. However, regions in the parameter space can be suitably selected, for example, by preparing the initial state with an appropriate value of the initial PM correlation such that the bounds for simultaneous and individual parameter estimation may coincide. In these cases, it remains possible to extract information about multiparameter models with precision comparable to that of a single-parameter estimation. For a detailed analysis of the behavior of the determinant of the QFIM, which is fundamental for attaining non-trivial precision of the quantum Cramér–Rao bound, see Appendix \ref{ap:detF}.

Figure~\ref{Fig5} depicts the dynamics of the derived bound,  $\widetilde{\mathcal{F}}_{\Lambda \Lambda}$, related to the estimation of the effective temperature parameter, $\Lambda(T)$, in the context of simultaneous multiparameter estimation. The analysis is carried out for several values of the initial correlation parameter $\gamma$, emphasizing its influence on the precision of the estimation process.  It is observed that the initial position-momentum correlation acts as an estimable resource, even within the framework of simultaneous multiparameter estimation. This behavior is consistent with previous findings in the single-parameter estimation regime, as reported in Ref.~\cite{Porto2024}. For the scenario of weak environmental interaction illustrated in Fig.~\ref{Fig5} (a), metrological performance improves with increasing absolute values of the parameter $\gamma$. In contrast, under strong environmental effects, states with negative values of $\gamma$ (\textit{contractive} states \cite{YuenPRL1983,Marinho_2020}) enable a more efficient parameter estimation compared to their anti-contractive counterparts (positive $\gamma$) or the standard uncorrelated Gaussian state ($\gamma = 0$).

To explain the behavior observed in Fig.~\ref{Fig5}, we analyze the asymptotic limits of $\widetilde{\mathcal{F}}_{\Lambda \Lambda}$ based on the QFIM elements derived in Appendix~\ref{appendix:QFIM}. In the regime of weak environmental effect (low-$\Lambda$), the bound expands as
\begin{equation}
    \Lambda^2 \widetilde{\mathcal{F}}_{\Lambda\Lambda}^{\text{low}} \approx 
    \frac{8 t^6 \Lambda^2 \hbar^2}{9 m^2 \tau_0^2} 
    + 
    \frac{2 t^2 \Lambda^2 \ell_0^4 \hbar^2 \left[ 3 \tau_0^2 + 3 \tau_0 t \gamma + t^2 (1+\gamma^2) \right]^2}{9 m^2 \tau_0^2 \sigma_0^2 (\ell_0^2 + \sigma_0^2)} .
\end{equation}
In this limit, the correlation parameter $\gamma$ appears within a squared polynomial $\left[ 3 \tau_0^2 + 3 \tau_0 t \gamma + t^2 (1+\gamma^2) \right]^2$, implying that even powers of $\gamma$ dominate. Consequently, the estimation precision is mainly sensitive to the magnitude $|\gamma|$, while the sign of the correlations is essentially irrelevant. Physically, because purity loss is minimal in this regime, the system dynamics are still dominated by coherent quantum spreading, and correlations enhance the Fisher information only at intermediate or long evolution times, when the system has enough time to build up sensitivity to the weak environmental parameter $\Lambda$. Conversely, under strong environmental effect (high-$\Lambda$), the asymptotic behavior becomes
\begin{equation}
    \Lambda^2 \widetilde{\mathcal{F}}_{\Lambda\Lambda}^{\text{high}} \approx 
    \frac{1}{2}
    +
    \frac{3 \tau_0^2}{42 \tau_0^2 + 48 \tau_0 t \gamma + 16 t^2 \gamma^2} .
\end{equation}
Here, the rapid loss of purity drives the system toward a classical statistical mixture. Correlations now enter linearly in the denominator, making the precision explicitly asymmetric and sensitive to the sign of $\gamma$. This result provides quantitative support for the advantage of \textit{contractive states} ($\gamma < 0$) \cite{YuenPRL1983,Marinho_2020}: their negative correlations partially counteract the broadening induced by environmental effects at short times, yielding higher precision than the uncorrelated or anti-contractive cases, before the noise eventually dominates.

\begin{figure*}[!ht]
\centering
\centering
\includegraphics[scale=0.28]{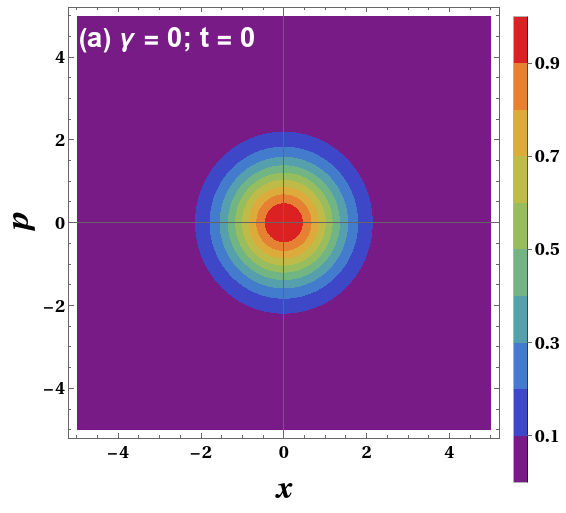}
\includegraphics[scale=0.28]{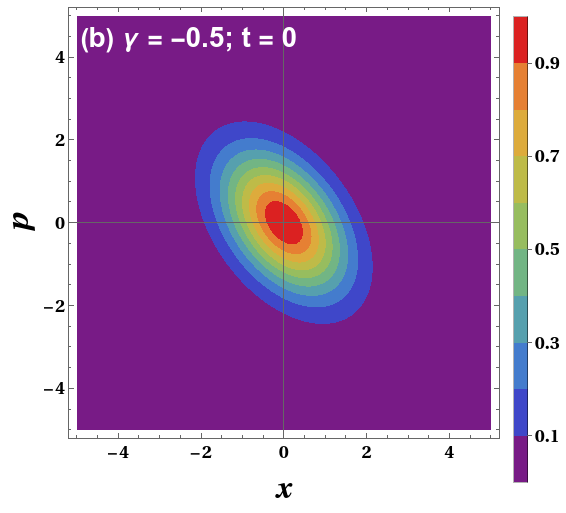}
\includegraphics[scale=0.28]{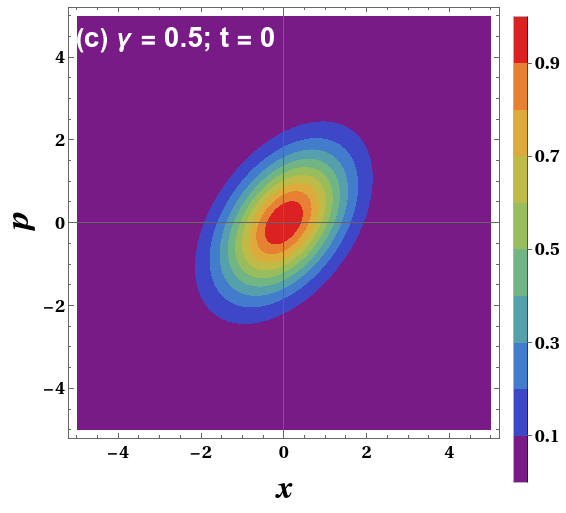}\\
\includegraphics[scale=0.28]{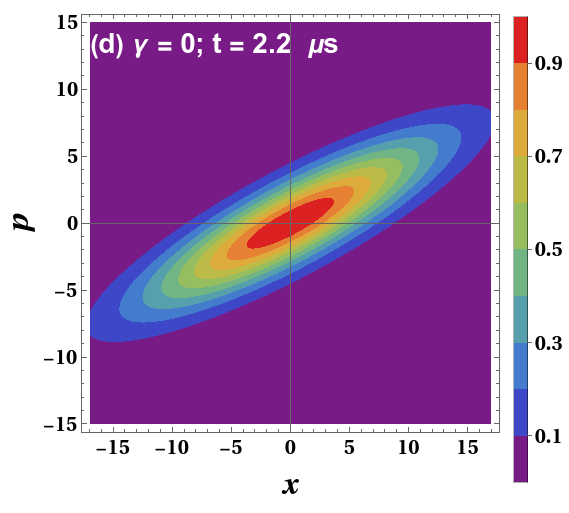}
\includegraphics[scale=0.28]{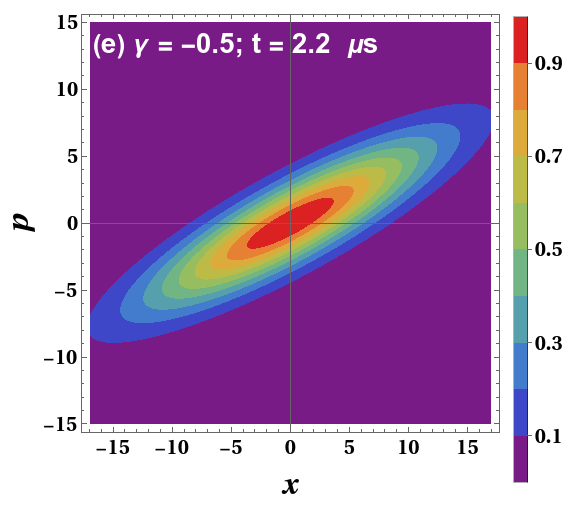}
\includegraphics[scale=0.28]{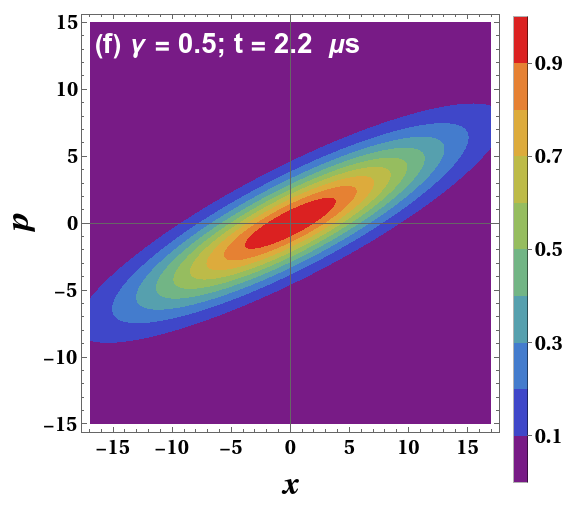}\\
\includegraphics[scale=0.28]{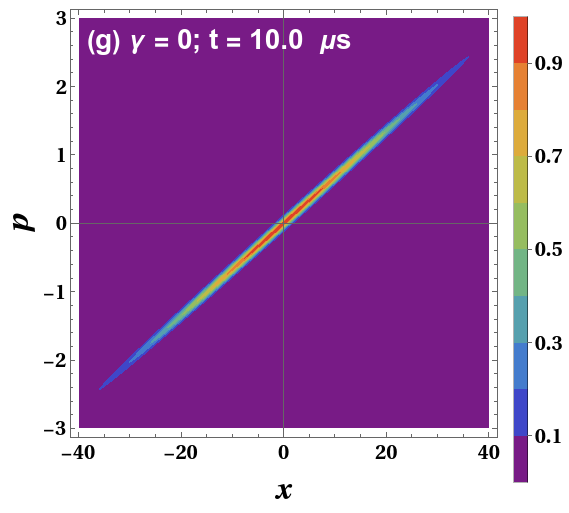}
\includegraphics[scale=0.28]{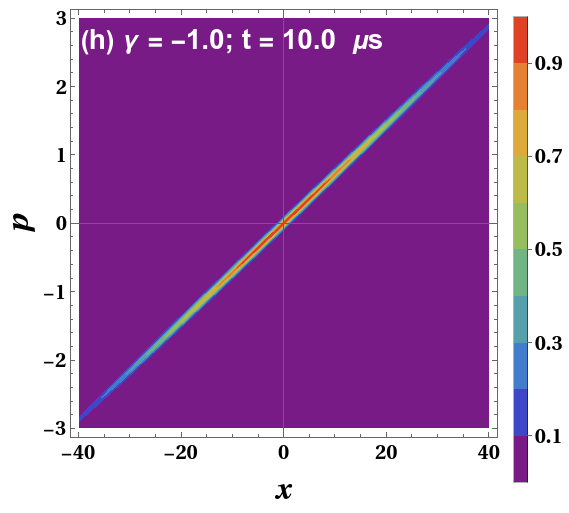}
\includegraphics[scale=0.28]{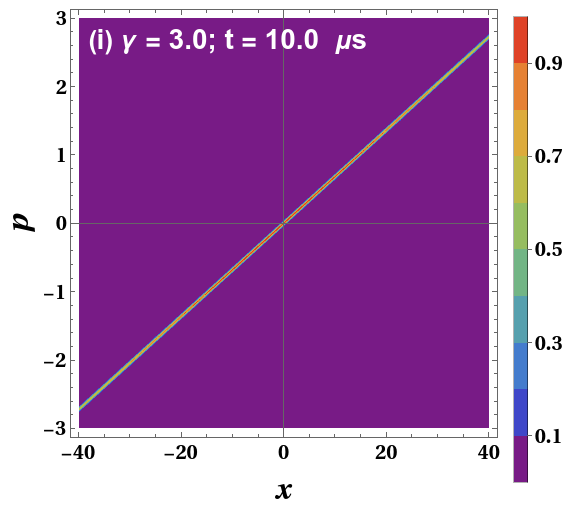}
\caption{The temporal evolution of the Wigner function under strong and weak environmental effects, characterized by $\Lambda = 3 \times 10^{22}\,\mathrm{m}^{-2}\mathrm{s}^{-1}$ ($T=786$ K) and $\Lambda = 3 \times 10^{15}\,\mathrm{m}^{-2}\mathrm{s}^{-1}$ ($T=16.9$ mK), respectively, is shown for different values of the initial correlation parameter $\gamma$. Panels (a), (d) and (g) correspond to the standard uncorrelated Gaussian state ($\gamma = 0$); panels (b), (e) and (h) represent the \textit{contractive} state ($\gamma < 0$); and panels (c), (f) and (i) correspond to the correlated state with $\gamma >0$. The top row illustrates the Wigner function at the initial time ($t = 0$); the middle row shows the evolved state at $t = 2.2\;\mu\mathrm{s}$ under strong environmental effects, while the bottom row shows the evolved state at $t = 10.0\;\mu\mathrm{s}$ under weak environmental effects. }\label{Fig6}
\end{figure*}

To provide insight into the observed metrological behavior of \textit{contractive} states and the distinct roles of correlation $\gamma$ in different regimes as pointed out in Fig.~\ref{Fig5}, we illustrate in Fig.~\ref{Fig6} the temporal evolution of the Wigner function under both strong and weak environmental effects. The evolution is shown for different values of the initial correlation parameter $\gamma$. Panels (a), (d) and (g) correspond to the standard uncorrelated Gaussian state ($\gamma = 0$); panels (b), (e) and (h) represent the \textit{contractive} state ($\gamma < 0$); and panels (c), (f) and (i) correspond to the correlated state with $\gamma >0$. The top row illustrates the Wigner function at the initial time ($t = 0$), the middle row shows the evolved state at $t = 2.2\;\mu\mathrm{s}$ under strong environmental effects, while the bottom row displays the evolved state at $t = 10.0\;\mu\mathrm{s}$ under weak environmental effects. In the latter case, a longer evolution time is selected because the system requires a more extended period to build up sensitivity to the perturbative environmental coupling.

In parameter estimation theory, the QFI can also be seen as a measure of the distinguishability between the quantum state $\rho_{\phi}$ and its neighboring state, $\rho_{\phi + \delta\phi}$~\cite{monras2013ARXIV, PinelPRA2013}. A larger change in the quantum state, corresponding to variations in the encoded parameter, implies higher sensitivity and, consequently, a greater quantum Fisher information. By examining the evolution of the Wigner function in the strong decoherence regime (middle row), we observe that although the final states appear qualitatively similar after the interaction with the Markovian channel, the \textit{contractive} state exhibits the greatest distinguishability. This is because it undergoes a $\sim90^\circ$ rotation to reach the elliptical shape oriented along $45^\circ$ associated with the evolved state. In contrast, a correlated initial state with $\gamma >0$ requires significantly less rotation. These differences in dynamics are interpreted as the source of the metrological advantage when using correlated states in the high-$\Lambda$ limit.

Conversely, in the weak environmental regime (bottom row), the dynamic is governed predominantly by coherent quantum spreading, and the loss of purity is minimal. Here, environmental effects act merely as a perturbation to the unitary evolution. As shown in panels (g), (h), and (i), the evolved Wigner functions exhibit phase-space deformations where the elliptical shape becomes more eccentric as the magnitude $|\gamma|$ increases. This is because the correlation parameter acts effectively as a squeezing source for the Wigner function~\cite{porto2025NonMarkovian,ThiagoNJP_2025,dearaujo2025gravimetryenhancementsqueezedstates}: while the vacuum state corresponds to $\gamma=0$, for a general squeezed state this parameter is given by $\gamma = \sinh(2r)\sin\phi$, where $r$ is the squeezing parameter and $\phi$ determines the phase-space orientation. Since the dominant dynamic is phase-space shearing (arising because points with higher momentum move faster, causing the distribution to shear sideways), the distinguishability depends primarily on the magnitude of the correlation $|\gamma|$ (and consequently $r$), rather than its sign (and consequently $\phi$). This confirms the behavior observed in Fig.~\ref{Fig5}(a), where correlations enhance the QFI but remain independent of the sign of $\gamma$. A similar analysis based on phase-space can be found in~\cite{Safranek_2015NJP,Safranek_2016PRA}.

\subsection{Compatibility condition}\label{sec:results_C}

Finally, to examine the existence of an optimal measurement strategy for the simultaneous estimation of the PM correlations and the effective coupling constant, we set $\phi_i = \gamma$ and $\phi_j = \Lambda$ in Eq.(\ref{eq:Safranek_2019JPA_Math_Theor}), employing the first and second statistical moments derived in Sec.~\ref{sec:theoretical_C}.  As $\boldsymbol{d} = 0$, this expression simplifies to
\begin{equation}
   \mathrm{Tr}(\rho [\mathcal{L}_{\gamma}, \mathcal{L}_{\Lambda}]) = \text{vec}[\partial_{\gamma}\boldsymbol{\sigma}]^{\text{T}} \mathcal{M}^{-1}(\boldsymbol{\sigma}\otimes \boldsymbol{\Omega}-\boldsymbol{\Omega}\otimes \boldsymbol{\sigma})\mathcal{M}^{-1}\text{vec}[\partial_{\Lambda}\boldsymbol{\sigma}].
\end{equation}
Using Eq.~(\ref{eq:Cov_matrix}), the matrix $\boldsymbol{\mathcal{M}}= \boldsymbol{\sigma} \otimes \boldsymbol{\sigma} - \boldsymbol{\Omega} \otimes \boldsymbol{\Omega}$ takes the form
\begin{equation}
    \boldsymbol{\mathcal{M}} =   \left ( \begin{array}{cccc}
\sigma_{xx}^2 \;\;& \sigma_{xx}\sigma_{xp}\;\;& \sigma_{xx}\sigma_{xp} \;\;& \sigma_{xp}^2 -1 \\
\sigma_{xx}\sigma_{xp} \;\;& \sigma_{xx}\sigma_{pp} \;\;&  \sigma_{xp}^2 +1 \;\; &  \sigma_{pp}\sigma_{xp}\\
\sigma_{xx}\sigma_{xp} \;\; & \sigma_{xp}^2 +1 \;\; &   \sigma_{xx}\sigma_{pp} \;\;&  \sigma_{pp}\sigma_{xp} \\
\sigma_{xp}^2 -1  \;\; & \sigma_{pp}\sigma_{xp}  \;\; &  \sigma_{pp}\sigma_{xp}   \;\; & \sigma_{pp}^2
\end{array} \right )\\,
\end{equation}
while the matrix $\boldsymbol{\sigma}\otimes \boldsymbol{\Omega}-\boldsymbol{\Omega}\otimes \boldsymbol{\sigma}$ simplifies to
\begin{equation}
\boldsymbol{\sigma}\otimes \boldsymbol{\Omega}-\boldsymbol{\Omega}\otimes \boldsymbol{\sigma} =   \left ( \begin{array}{cccc}
0 \;\;& \sigma_{xx}\;\;& -\sigma_{xx} \;\;& 0 \\
-\sigma_{xx} \;\;& 0 \;\;&  -2\sigma_{xp} \;\; &  -\sigma_{pp}\\
\sigma_{xx} \;\; & 2\sigma_{xp} \;\; &   0 \;\;&  \sigma_{pp} \\
0  \;\; & \sigma_{pp}  \;\; &  -\sigma_{pp}   \;\; & 0
\end{array} \right )\\.
\end{equation}
Therefore, the product $\boldsymbol{\mathcal{M}}^{-1}(\boldsymbol{\sigma}\otimes \boldsymbol{\Omega}-\boldsymbol{\Omega}\otimes \boldsymbol{\sigma})\boldsymbol{\mathcal{M}}^{-1}$ can be written as
\begin{equation}
\boldsymbol{\mathcal{M}}^{-1}(\boldsymbol{\sigma}\otimes \boldsymbol{\Omega}-\boldsymbol{\Omega}\otimes \boldsymbol{\sigma})\boldsymbol{\mathcal{M}}^{-1} = \frac{1}{[1-\text{det}(\boldsymbol{\sigma})]^2}  \left ( \begin{array}{cccc}
0 \;\;& \sigma_{pp} \;\;& -\sigma_{pp}  \;\;& 0 \\
 -\sigma_{pp}  \;\;& 0 \;\;&  2\sigma_{xp}   \;\; &  -\sigma_{xx}   \\
 \sigma_{pp}  \;\; &  -2\sigma_{xp}  \;\; & 0 \;\;&  \sigma_{xx}   \\
 0 \;\; &  \sigma_{xx}   \;\; &  -\sigma_{xx}  \;\; & 0
\end{array} \right )\\.
\end{equation}
Furthermore, since we are dealing with an open quantum system, no inconsistency arises under these conditions, as $\text{det}(\boldsymbol{\sigma}) \neq 1$. Finally, by multiplying both sides by the corresponding vectorized matrices, we obtain
\begin{gather}
\mathrm{Tr}[(\rho [\mathcal{L}_{\gamma}, \mathcal{L}_{\Lambda}]) =\text{vec}[\partial_{\gamma}\boldsymbol{\sigma}]^{\text{T}} \mathcal{M}^{-1}(\boldsymbol{\sigma}\otimes \boldsymbol{\Omega}-\boldsymbol{\Omega}\otimes \boldsymbol{\sigma})\mathcal{M}^{-1}\text{vec}[\partial_{\Lambda}\boldsymbol{\sigma}] = \frac{1}{[1-\text{det}(\boldsymbol{\sigma})]^2} \nonumber\\
\left (\begin{array}{cccc}
  \partial_{\gamma}\sigma_{xx}   & \partial_{\gamma}\sigma_{xp} &  \partial_{\gamma}\sigma_{xp}   & \partial_{\gamma}\sigma_{pp} \\
\end{array}\right )  \left ( \begin{array}{cccc}
0 \;\;& \sigma_{pp} \;\;& -\sigma_{pp}  \;\;& 0 \\
 -\sigma_{pp}  \;\;& 0 \;\;&  2\sigma_{xp}   \;\; &  -\sigma_{xx}   \\
 \sigma_{pp}  \;\; &  -2\sigma_{xp}  \;\; & 0 \;\;&  \sigma_{xx}   \\
 0 \;\; &  \sigma_{xx}   \;\; &  -\sigma_{xx}  \;\; & 0
\end{array} \right )  \left (\begin{array}{cccc}
  \partial_{\Lambda}\sigma_{xx}\\ \partial_{\Lambda}\sigma_{xp}\\  \partial_{\Lambda}\sigma_{xp}\\ \partial_{\Lambda}\sigma_{pp} \\
\end{array}\right ) = 0.  \label{eq:compatibility}
\end{gather}

This demonstrates, in principle, the possibility of saturating the quantum Cramér–Rao bound, as well as the attainability of the bounds $\widetilde{\mathcal{F}}_{\gamma\gamma}$ and $\widetilde{\mathcal{F}}_{\Lambda\Lambda}$ derived in the present work. For example, considering the explicit expressions for the covariance matrix derived in Eqs.~(\ref{eq:sigmaxx_sigampp}) and (\ref{eq:sigmaxp}), it is straightforward to show that
\begin{gather}
    \partial_{\gamma}\sigma_{xx}= \frac{2t(\gamma t +\tau_0)}{\tau_0^2}   ,\;\;\;\;  \partial_{\gamma}\sigma_{xp}= 1+\frac{2\gamma t}{\tau_0}  , \;\;\;\;\ \partial_{\gamma}\sigma_{pp}=2\gamma \  , \nonumber\\
\partial_{\Lambda}\sigma_{xx} =\frac{4t^3\sigma_0^2}{3\tau_0^3} , \;\;\;\;  \partial_{\Lambda}\sigma_{xp}= \frac{2t^2\sigma_0^2}{\tau_0}        , \;\;\;\; \partial_{\Lambda}\sigma_{pp} = 4t\sigma_0^2.
\end{gather}
By substituting these explicit derivatives into Eq.~(\ref{eq:compatibility}), one can readily verify the compatibility condition.

\section{Discussion}\label{sec:disc}

In this work, we explored the role of position–momentum (PM) correlations as a quantum resource for multiparameter estimation within the framework of Gaussian states. We introduced a model that incorporates both unitary and non-unitary dynamics to account for realistic open-system effects commonly encountered in quantum sensing. Our results show that the use of PM-correlated probe states can improve precision bounds in simultaneous parameter estimation, outperforming individual estimation approaches within specific and limited parameter ranges. Furthermore, we demonstrate that the quantum Cramér–Rao bound is saturable for the two key parameters studied, PM correlation and effective temperature, highlighting the practical relevance of our method.

By analyzing the evolution of the quantum Fisher information matrix (QFIM) under a scattering-induced decoherence channel, a dominant source of decoherence in physical systems~\cite{Schlosshauer2,joos2003decoherence}, we demonstrated that entanglement with environmental degrees of freedom causes spatial decoherence, yet does not eliminate the metrological advantage provided by PM correlations. Our model assumes the long-wavelength limit~\cite{Schlosshauer2}, which ensures weak system–environment coupling and supports effective Markovian dynamics. The environment is modeled as a gas of particles obeying a Maxwell–Boltzmann distribution at temperatures well above the ultracold regime~\cite{CornellScienceBSC1995}.

 While the present work focuses on pure decoherence under the Born-Markov approximation, future studies could extend this framework by incorporating dissipative processes alongside decoherence~\cite{Schlosshauer2,PP2024} and accounting for non-Markovian memory effects~\cite{FNori2015SciRep}. Such extensions would advance the description of open quantum system dynamics beyond the Born-Markov paradigm~\cite{AnJunHongPRApplied2022,BreuerPRA2020,porto2025NonMarkovian}, ultimately enabling a more complete understanding of metrology in quantum environments. Furthermore, extending the model to include more molecular degrees of freedom, such as rotational and vibrational modes, could significantly enhance the sensitivity of molecular sensors acting as quantum probes~\cite{Hutzler_2020QuanScTech,DeMille2024Nature}, therefore broadening their impact in emerging quantum technologies.

To this end, our findings demonstrate that metrological enhancements are achievable through correlations between conjugate observables, without relying on entanglement, an experimentally demanding resource. This opens promising new pathways in quantum metrology, emphasizing alternative quantum resources that may be more accessible for practical applications.

\begin{acknowledgments}
J. C. P. Porto acknowledges Fundação de Amparo à Pesquisa do Estado do Piauí (FAPEPI) for financial support. C.H.S.V acknowledges the São Paulo Research Foundation (FAPESP), Grant. No. 2023/13362-0, for financial support and the Federal University of ABC (UFABC) to provide the workspace. P.R.D acknowledges support from the NCN Poland, ChistEra-2023/05/Y/ST2/00005 under the project Modern Device Independent Cryptography (MoDIC). I.G.P. acknowledges Grant No. 306528/2023-1 from CNPq and L.S.M. acknowledges support from the National Institute of Science and Technology on National Institute of Photonics (INFO) CNPq - INCT grant 409174/2024-6.
\end{acknowledgments}

\section*{Availability of Data and Material}

All data generated or analyzed during this study are included in this published article.

\appendix

\section{Parameters of density matrix for a correlated Gaussian state and purity}\label{appendix:rho_parameters}

The explicit expressions for the parameters of the time-evolved density matrix, presented in Eq.~(\ref{rho_Lambda}), are obtained by solving the dynamics of the open quantum system governed by the scattering decoherence model. By integrating the initial correlated Gaussian state with the time-evolution propagator $K_{\Lambda}(x,x',t;x_0,x'_0,0)$ (as detailed in Ref.~\cite{Porto2024}), and utilizing the standard Gaussian integral identity $\int_{-\infty}^{\infty} e^{-ax^2+bx} dx = \sqrt{\frac{\pi}{a}} e^{\frac{b^2}{4a}}$, the resulting coefficients are derived as follows:

\begin{gather}
\mathcal{A}_{t}=A_{1}+A_{2}-iA_{3}, \quad     \mathcal{B}_{t}=A_{1}-A_{2}-iA_{3}, \quad \mathcal{C}_{t}=2iA_{3},
\end{gather}
\begin{gather}
A_{1} = \frac{m^{2}}{8 \hbar^2t^2 \sigma_0^2 B^2 }, \quad
B^2 =  \frac{1}{4\sigma_0^4} + \frac{1}{2\ell_0^2\sigma_0^2} +\left( \frac{m}{2 \hbar t} + \frac{\gamma}{2\sigma_0^2} \right)^2  + \frac{\Lambda t}{3\sigma_0^2},
\end{gather}
\begin{gather}
A_2 = \frac{m^{2}}{4 \hbar^2t^2 B^2} \left( \frac{1}{2 \ell_0^2} + \Lambda t \right) + \frac{\Lambda t}{12 \sigma_0^2 B^2} \left( \Lambda t + \frac{1}{2\sigma_0^2} + \frac{2}{\ell_0^2} \right) 
+ \frac{m \Lambda \gamma}{4 \hbar \sigma_0^2 B^2} + \frac{\Lambda t \gamma^2}{12 \sigma_0^4 B^2},
\end{gather}
and
\begin{align}
A_3 = \frac{m}{4 \hbar t \sigma_0^2 B^2} \left( \Lambda t + \frac{1}{2\sigma_0^2} + \frac{1}{\ell_0^2} \right) + \frac{m \gamma}{8 \hbar t \sigma_0^2 B^2}\left( \frac{m}{ \hbar t} + \frac{\gamma}{\sigma_0^2} \right),  \;\;\;\;\;\;\; \mathcal{N}_t = \sqrt{\frac{2A_{1}}{\pi}}.
\end{align}

\section{Expressions for the Elements of the Quantum Fisher Information Matrix}\label{appendix:QFIM}

In this Appendix, we present the explicit analytical expressions for the elements of the Quantum Fisher Information Matrix (QFIM). These results are derived by substituting the covariance matrix elements $\sigma_{xx}$, $\sigma_{pp}$, and $\sigma_{xp}$ (presented in Eqs.~(\ref{eq:Cov_matrix}), (\ref{eq:sigmaxx_sigampp}), and (\ref{eq:sigmaxp}) of the main text) and their respective derivatives into the general definition given by Eq.~(\ref{eq:Dominik}). The resulting matrix elements are:

\begin{gather}
\mathcal{F}_{\gamma \gamma}= (2 \alpha)^{-1} \ell_{0}^{2} \{   \left( 12 m^{2} \sigma_{0}^{4} \ell_{0}^{2}\hbar^{2}t^{4} + 16 \ell_{0}^{2} \hbar^{4}  \gamma^{2} t^{6} + 48 m \sigma_{0}^{2} \ell_{0}^{2} \hbar^{3}  \gamma t^{5} \right) \Lambda^{2}  +  \nonumber \\ + 
 \left(  6 m^{2} \sigma_{0}^{2} \hbar^{2}t^{3}  \left(2\sigma_{0}^{2} + \beta \ell_{0}^{2} \right) + 18 m^{4} \sigma_{0}^{6}   \ell_{0}^{2} t + 18 m^{3} \sigma_{0}^{4} \ell_{0}^{2} \hbar  \gamma t^{2}  \right) \Lambda + 9 m^{4} \sigma_{0}^{6} \},
\end{gather}
\begin{gather}
\mathcal{F}_{\gamma \Lambda} = \alpha^{-1} \{\ell_{0}^{2} \sigma_{0}^{2} \hbar t \left[ 2 \ell_{0}^{2} \hbar^{2} t^{4} \left( 3 m \sigma_{0}^{2} + 2 \gamma t \hbar \right) \Lambda^{2} - 3 m^{2} \sigma_{0}^{2} \left( 3 m \sigma_{0}^{2} + 2 \gamma t \hbar \right) \right] \},
\end{gather}
\begin{gather}
\mathcal{F}_{\Lambda \Lambda}= \alpha^{-1} \{ 2  t^{2} \left[ 2 \sigma_{0}^{4} \ell_{0}^{4} t^{2} \Lambda^{2} +2 \sigma_{0}^{2} \ell_{0}^{2} \left( 2 \sigma_{0}^{2} t^{5} \hbar^{4} + \ell_{0}^{2} \beta t^{5} \hbar^{4} + 3m \sigma_{0}^{2} \ell_{0}^{2} \gamma t^{4} \hbar^{3} + 3m^{2} \sigma_{0}^{4} \ell_{0}^{2} t^{3} \hbar^{2} \right) \Lambda  \right] \nonumber + \\
 +\left[ 6 m \sigma_{0}^{2} \ell_{0}^{2} \gamma t^{3} \hbar^{3} \left( 2\sigma_{0}^{2} + \ell_{0}^{2} \beta  \right) + \hbar^{4} t^{4} \beta \left( \ell_{0}^{4} \beta + 4 \sigma_{0}^{2} \ell_{0}^{2}  \right) +9 m^{4} \sigma_{0}^{8} \ell_{0}^{4} + 18 m^{3}\sigma_{0}^{6} \ell_{0}^{4} \gamma t \hbar  \right] \}  ,
\end{gather}
where
\begin{align}
\alpha = \{  2 \sigma_{0}^{2} \ell_{0}^{2} t^{4} \hbar^{2} \Lambda^{2} + \left( 6 m \sigma_{0}^{2} \ell_{0}^{2} \hbar \gamma t^{2} + 2 t^{3} \hbar^{2} \left( 2 \sigma_{0}^{2} + \beta \ell_{0}^{2}  \right) + 6 m^{2} \sigma_{0}^{4} \ell_{0}^{2} t  
 \right) \Lambda + 3 m \sigma_{0}^{4} \} \left(1+\frac{\ell_{0}^{2}}{\sigma_{0}^{2}} \right) 
\end{align}
and 
\begin{align}
\beta = 1 + \gamma^{2}
\end{align}

\section{Determinant of the Quantum Fisher Information Matrix (QFIM)}\label{ap:detF}

\begin{figure*}[!ht]
\centering
\includegraphics[scale= 0.4]{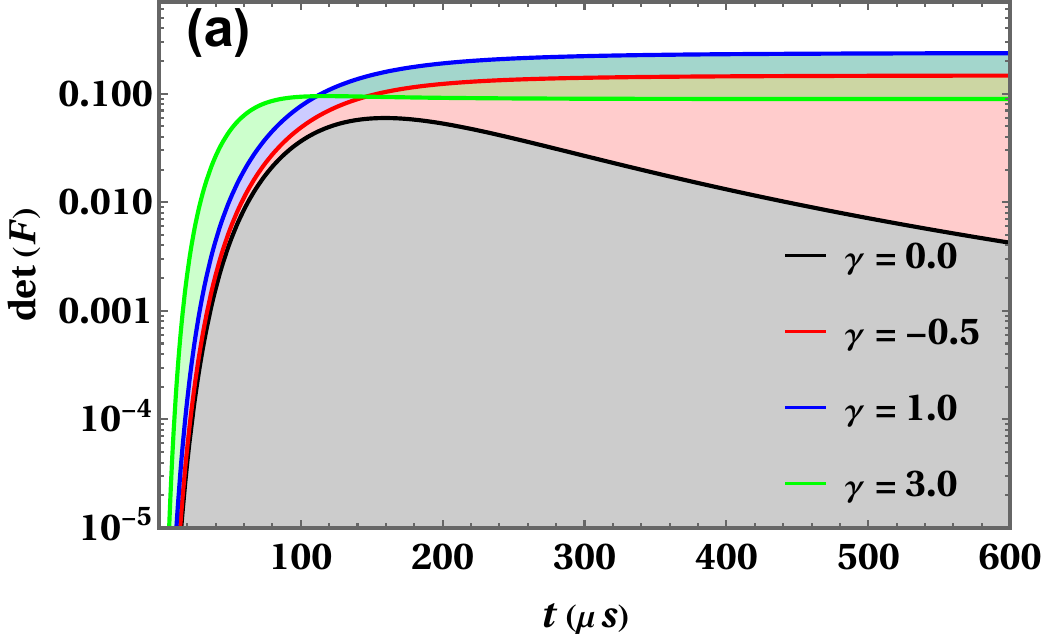}
\includegraphics[scale= 0.4]{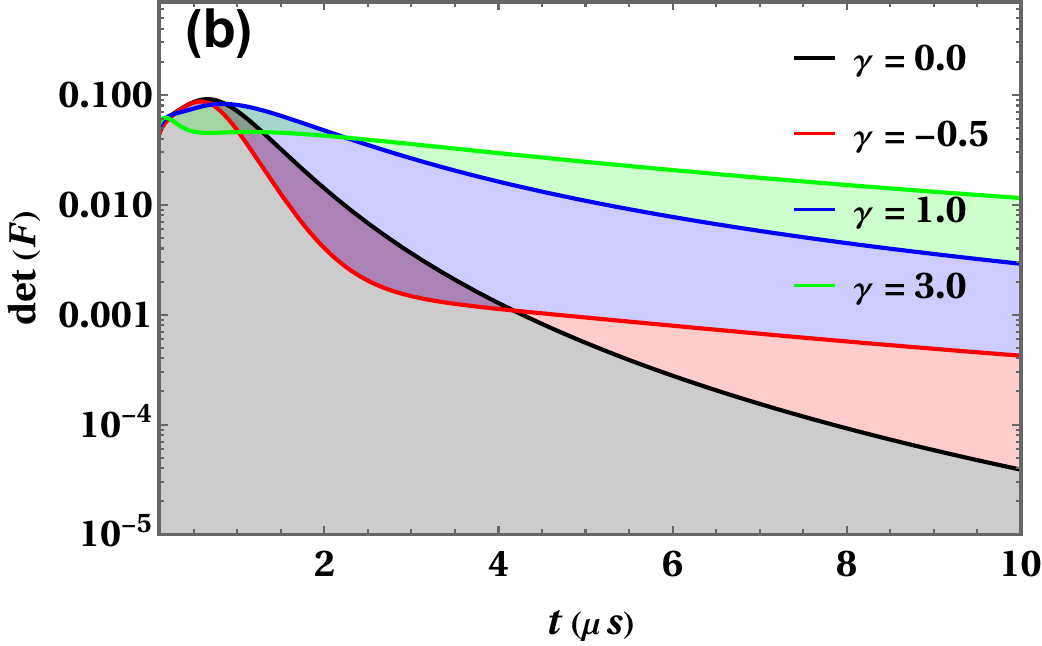}
\caption{Temporal dynamics of the determinant of the Quantum Fisher Information Matrix (QFIM) for different values of the initial correlation parameter $\gamma$. (a) corresponds to a weak environmental effect characterized by $\Lambda = 3 \times 10^{15}\mathrm{m}^{-2}\mathrm{s}^{-1}$ ($T=16.9$ mK), while (b) represents a strong environmental effect with $\Lambda = 3 \times 10^{22}\mathrm{m}^{-2}\mathrm{s}^{-1}$ ($T=786$ K).}\label{Fig7}
\end{figure*}

Since the quantum Cramér–Rao bound is expressed as a matrix inequality that holds only when the Quantum Fisher Information Matrix (QFIM) is invertible, it is crucial to analyze the behavior of the determinant of the QFIM. This quantity is fundamental for attaining non-trivial precision in the context of multiparameter estimation. A singular QFIM, characterized by $\det(\boldsymbol{\mathcal{F}}) = 0$, typically indicates that the unknown parameters are not all independent and therefore cannot be estimated simultaneously. In Fig.~\ref{Fig7}, we present the temporal dynamics of the determinant of the QFIM for different values of the initial correlation parameter $\gamma$, under both weak and strong environmental effects. These results demonstrate conditions under which both parameters remain estimable simultaneously.

\FloatBarrier % Forces all floats before this point
%\clearpage % Ensures that all figures are processed before the references

\bibliography{references}% Produces the bibliography via BibTeX.

\end{document}